\documentclass[seceq]{ptptex}
\usepackage{wrapft}
\usepackage{graphicx}% Include figure files
\usepackage{epsfig}

\def\Im{{\rm Im}}
\def\Re{{\rm Re}}
\renewcommand\epsilon{\varepsilon}

\newcommand{\ov}{\overline}
\newcommand{\f}{\frac}

\newcommand{\nn}{\nonumber}
\newcommand{\m}{\mbox}

\def\nn{\nonumber}
\def\bea{\begin{eqnarray}}
\def\eea{\end{eqnarray}}
\def\ynu{y_{\nu}}

\def\lpN{l^+_i \phi^- \rightarrow N_{R_k}(s_z)}
\def\lmN{l^-_i \phi^+ \rightarrow N_{R_k}(s_z)}
\def\nN{\nu_i \phi^{0}\rightarrow N_{R_k}(s_z)}
\def\ansN{\bar{\nu}_i \phi^{0*} \rightarrow N_{R_k}(s_z)}
\def\Nlp{N_{R_k}(s_z) \rightarrow l^+_i \phi^-}
\def\Nlm{N_{R_k}(s_z)\rightarrow l^-_i \phi^+}
\def\Nans{N_{R_k}(s_z) \rightarrow \bar{\nu}_i \phi^{0*}}
\def\Nn{N_{R_k}(s_z) \rightarrow\nu_i \phi^{0}}
\def\lplpij{l^+_i\phi^- \rightarrow l^+_j \phi^-}
\def\lplpji{l^+_j \phi^- \rightarrow l^+_i \phi^-}
\def\lpnij{l^+_i\phi^- \rightarrow \bar{\nu}_j \phi^{0*}}
\def\anlpji{\bar{\nu}_j \phi^{0*}\rightarrow l_i^+ \phi^-}
\def\annij{\bar{\nu}_i \phi^{0*}\rightarrow \bar{\nu}_j \phi^{0*}}
\def\annji{\bar{\nu}_j \phi^{0*}\rightarrow\bar{\nu}_i \phi^{0*}}
\def\nlmji{\nu_j \phi^0\rightarrow l_i^- \phi^+}
\def\lmnij{l^-_i\phi^+ \rightarrow \nu_j \phi^0}
\def\lmlmji{l^-_j\phi^+ \rightarrow l^-_i \phi^+}
\def\lmlmij{l^-_i \phi^+ \rightarrow l^-_j \phi^+}
\def\lpnji{l^+_j\phi^- \rightarrow \bar{\nu}_i \phi^{0*}}
\def\anlpij{\bar{\nu}_i \phi^{0*}\rightarrow l_j^+ \phi^-}
\def\nnij{\nu_i \phi^0\rightarrow \nu_j \phi^0}
\def\nnji{\nu_j \phi^0\rightarrow \nu_i \phi^0}
\def\nlmij{\nu_i \phi^0\rightarrow l_j^- \phi^+}
\def\lmnji{l_j^- \phi^+\rightarrow \nu_i \phi^0}
\def\lplmi{l^+_i\phi^- \rightarrow l^-_j \phi^+}

\def\lpni{l^+_i\phi^- \rightarrow \nu_j \phi^0}
\def\lmani{l^-_j\phi^+ \rightarrow \bar{\nu}_i \phi^{0*}}
\def\anni{\bar{\nu}_i \phi^{0*}\rightarrow \nu_j \phi^0}
\def\anlmi{\bar{\nu}_j \phi^{0*}\rightarrow l_i^- \phi^+}
\def\nani{\nu_j \phi^0\rightarrow \bar{\nu}_i \phi^{0*}}
\def\nlpi{\nu_j \phi^0\rightarrow l_i^+ \phi^-}
\def\pni{\phi^0\phi^0\rightarrow \bar{\nu_i}\bar{\nu}_j}
\def\ninj{\nu_i\nu_j\to\phi^{0*}\phi^{0*}}
\def\pmpzsi{\phi^-\phi^{0*}\rightarrow l^-_i\nu_j}
\def\lmni{l^-_i\nu_j\to \phi^-\phi^{0*}}
\def\pppzi{\phi^+\phi^0\rightarrow l^+_i\bar{\nu}_j}
\def\lpani{ l^+_i\bar{\nu}_j\to \phi^+\phi^0}
\def\psani{\phi^{0*}\phi^{0*}\rightarrow \nu_i\nu_j}
\def\anani{\bar{\nu_i}\bar{\nu}_j\to\phi^0\phi^0}
\def\pmlmi{\phi^-\phi^-\rightarrow l_i^-l_j^-}
\def\pplpi{\phi^+\phi^+\rightarrow l_i^+l_j^+}
\def\lmlmi{l_i^-l_j^-\to\phi^-\phi^-}
\def\lplpi{l_i^+l_j^+\to\phi^+\phi^+}
\def\lplmj{l^+_j\phi^- \rightarrow l^-_i \phi^+}
\def\lmlpj{l^-_i \phi^+ \rightarrow l^+_j \phi^-}
\def\lpnj{l^+_j\phi^- \rightarrow \nu_i \phi^0}
\def\lmanj{l^-_i\phi^+ \rightarrow \bar{\nu}_j \phi^{0*}}
\def\annj{\bar{\nu}_j \phi^{0*}\rightarrow \nu_i \phi^0}
\def\anlmj{\bar{\nu}_i \phi^{0*}\rightarrow l_j^- \phi^+}
\def\nanj{\nu_i \phi^0\rightarrow \bar{\nu}_j \phi^{0*}}
\def\nlpj{\nu_i \phi^0\rightarrow l_j^+ \phi^-}
\def\pmpzsj{\phi^-\phi^{0*}\rightarrow l^-_j\nu_i}
\def\lmnj{l^-_j\nu_i\to \phi^-\phi^{0*}}
\def\pppzj{\phi^+\phi^0\rightarrow l^+_j\bar{\nu}_i}
\def\lpanj{ l^+_j\bar{\nu}_i\to \phi^+\phi^0}
\def\Ntlmb{N_{R_k}(s_z)\bar{t}_L^\alpha \rightarrow l_i^-\bar{b}_R^\alpha}
\def\lmb{l_i^-\bar{b}_L^\alpha\to N_{R_k}(s_z)\bar{t}_R^\alpha}
\def\Ntlpb{N_{R_k}(s_z)t_R^\alpha\rightarrow l_i^+b_L^\alpha}
\def\lpb{l_i^+b_L^\alpha\to N_{R_k}(s_z)t_R^\alpha}
\def\Nblpt{N_{R_k}(s_z)\bar{b}_L^\alpha\rightarrow l_i^+\bar{t}_R^\alpha}
\def\lpt{l_i^+\bar{t}_R^\alpha\to N_{R_k}(s_z)\bar{b}_L^\alpha}
\def\Nblmt{N_{R_k}(s_z)b_L^\alpha\rightarrow l_i^-t_R^\alpha}
\def\lmt{l_i^-t_R^\alpha\to N_{R_k}(s_z)b_L^\alpha}

\def\Nlptb{N_{R_k}(s_z)l_i^+\to\bar{b}_L^\alpha t_R^\alpha}
\def\Nlmtb{N_{R_k}(s_z)l_i^-\to b_L^\alpha\bar{t}_R^\alpha}
\def\tbNlp{\bar{b}_L^\alpha t_R^\alpha\to N_{R_k}(s_z)l_i^+}
\def\tbNlm{b_L^\alpha\bar{t}_R^\alpha\to N_{R_k}(s_z)l_i^-}
\def\Nntt{N_{R_k}(s_z)\nu_i\to t_L^\alpha\bar{t}_R^\alpha}
\def\Nantt{N_{R_k}(s_z)\bar{\nu_i}\to \bar{t}_L^\alpha t_R^\alpha}
\def\Ntnt{N_{R_k}(s_z)t_L^\alpha\to t_R^{\alpha}\nu_i}
\def\Ntant{N_{R_k}(s_z)t_R^\alpha\to t_L^\alpha{\bar\nu}_i}
\def\Natnat{N_{R_k}(s_z)\bar{t}_R^\alpha\to \bar{t}_L^\alpha\nu_i}
\def\Natanat{N_{R_k}(s_z)\bar{t}_L^\alpha\to \bar{t}_R^{\alpha}\bar{\nu}_i}
\def\ttNn{t_L^\alpha\bar{t}_R^\alpha\to N_{R_k}(s_z)\nu_i}
\def\ttNan{t_R^\alpha\bar{t}_L^\alpha\to N_{R_k}(s_z)\bar{\nu_i}}
\def\ntNt{t_R^\alpha\nu_i\to N_{R_k}(s_z)t_L^\alpha}
\def\antNt{t_L^\alpha{\bar\nu}_i\to N_{R_k}(s_z)t_R^\alpha}
\def\natNat{\bar{t}_L^\alpha\nu_i\to N_{R_k}(s_z)\bar{t}_R^\alpha}
\def\anatNat{\bar{t}_R^\alpha\bar{\nu}_i\to N_{R_k}(s_z)\bar{t}_L^\alpha}
\def\flmi{f_{l_i^-}}
\def\flpi{f_{l_i^+}}
\def\fni{f_{\nu_i}}
\def\fani{f_{\bar{\nu}_i}}
\def\flmj{f_{l_j^-}}
\def\flpj{f_{l_j^+}}
\def\fnj{f_{\nu_j}}
\def\fanj{f_{\bar{\nu}_j}}
\def\fpp{f_{\phi^+}}
\def\fpm{f_{\phi^-}}
\def\fpz{f_{\phi^0}}
\def\fpzs{f_{\phi^{0*}}}
\title{Primordial Lepton Family Asymmetries in Seesaw Model}
\author{ Tomohiro \textsc{Endoh}\footnote{E-mail: endoh@theo.phys.sci.hiroshima-u.ac.jp},
          Takuya
          \textsc{Morozumi}\footnote{E-mail:morozumi@theo.phys.sci.hiroshima-u.ac.jp},
          and Zhaohua  \textsc{Xiong}\footnote{E-mail:
          xiongzh@theo.phys.sci.hiroshima-u.ac.jp}
}
\inst{Graduate School of Science, Hiroshima University, \\
Higashi Hiroshima 739-8526, Japan}

\abst{In leptogenesis scenario, the decays of heavy Majorana neutrinos 
generate 
lepton family asymmetries, $Y_e , Y_{\mu}$ and $Y_{\tau}$.
They are sensitive to CP violating
phases in seesaw models. 
The time evolution of
the lepton family asymmetries
are derived by solving Boltzmann equations.
By taking a minimal seesaw model,
we show how each family asymmetry varies with a CP violating phase.
For instance, we find the case that the lepton asymmetry is dominated
by $Y_{\mu}$ or $Y_{\tau}$ depending on the choice of the
 CP violating phase.
We also find the case that
the signs of lepton family asymmetries 
$Y_{\mu}$ and $Y_{\tau}$ are opposite each other.
Their absolute values can be larger than the total lepton asymmetry and 
the baryon asymmetry may
result from the cancellation of the lepton family asymmetries.}
\begin{document}
\maketitle

\section{Introduction}
From the recent astronomical observations,
the precise values
for cosmological baryon to photon ratio 
($\eta=\frac{n_B}{n_{\gamma}}$) are obtained,
\bea
\eta=&&(6.5\pm^{0.4}_{0.3}){\times}10^{-10}~~
\mbox{~\cite{baryon-photon}}, \quad (6.0\pm^{1.1}_{0.8}){\times}10^{-10}~~
\mbox{~\cite{BOOMERanG}}.
\label{eta}
\eea
Leptogenesis~\cite{FuYa} is a very attractive scenario for
the matter
and the anti-matter asymmetry of our universe. \cite{Yoshimura}  
Because the seesaw mechanism~\cite{seesaw} which underlies the scenario is able to give a
natural explanation for tiny neutrino masses, 
it is important to study the scenario from both cosmological
and particle physics point of views. 
In the early literatures~\cite{Plumacher,leptogenesis-1},
the lepton number production and the evolution in the expanding universe
were studied. In the last few years, much
attention has been paid to CP violation of seesaw
models. The relation between CP violation of 
leptogenesis and CP violation of 
neutrino oscillation was 
shown~\cite{correlation}.  Because there are many
CP violating phases in seesaw models, it is important to identify CP
violating observables as much as possible.
Such observables include the low energy CP
asymmetry of neutrino oscillations. In the present paper, we study 
the lepton family density to entropy density ratios   
$Y_{e},~Y_{\mu}$ and $~Y_{\tau}$ in seesaw models. 
Hereafter, we call them
the lepton family asymmetries. We shall show that they
are sensitive to the CP violating phases.
By studying the lepton family asymmetries, we can trace more detailed 
history of the leptogenesis scenario.

So far, lack of our knowledge for the neutrino Yukawa couplings in
seesaw models has resulted in numerous speculations on their possible
forms. Our result shows that different forms may lead to different
lepton family asymmetries.  Thus, the lepton family asymmetries are
useful for constraining the forms of the Yukawa couplings. 

In this paper, we give a detailed calculation for the lepton family
asymmetries by solving the Boltzmann equations.
In the Boltzmann equations of the previous works,~\cite{Plumacher}
~\cite{leptogenesis-1} it is implicitly assumed that
the chemical potentials are the same for all
the three lepton doublets and the effects of the 
chemical potentials of the other particles are neglected.
Contrastingly, we shall consider all the chemical potentials 
for the standard
model (SM) particles.~\cite{Spectator}.
We do not assume that the three lepton doublets have
the same chemical potential a priori. 
We assign the independent chemical potentials for each generation of lepton 
doublets.~\cite{Barbieri}. In this way, we can treat the evolution
of the lepton family asymmetries in the Boltzmann equation.

For numerical analysis, we extend the previous study \cite{our-paper} 
on the {\em minimal} seesaw model. In the model, 
there are three CP violating phases.
Considering the lepton family asymmetries,
we find that they are sensitive  
to one of the CP violating phases which is different from the 
one related to the total lepton asymmetry $Y_L$.  For instance,
we find the case that
the signs of the lepton family asymmetries $Y_{\mu}$ and $Y_{\tau}$
are opposite each other and the tiny 
lepton asymmetry $Y_L$ results from the
cancellation among the much larger family asymmetries.

The paper is organized as follows. In Sec. \ref{sec:model}, 
the Yukawa sector for seesaw models is given and 
the $minimal$ seesaw model is introduced.
In Sec. \ref{sec:be},
The Boltzmann equations for the lepton family asymmetries are derived.
Sphaleron process and the other equilibrium processes are discussed
and the relations among 
chemical potentials are studied.
Using the minimal seesaw model and
the relations of the
chemical potentials, we solve the Boltzmann equations
and present the numerical results in Sec. \ref{sec:nr}.
Finally, in Sec. \ref{sec:con}, we give our conclusions. 
The derivations of various cross sections used in our calculation are 
shown in Appendices.

%%%%%%%%%%%%%%%%%%%%%%%%%%%%%%%%%%%%%%%%%%%%%%%%%%%
%%%%%%%%%%%  Model  %%%%%%%%%%%%%%%%%%%%%%%%%%%%%%%%
%%%%%%%%%%%%%%%%%%%%%%%%%%%%%%%%%%%%%%%%%%%%%%%%%%% 
\section{Model}
\label{sec:model}

We first discuss the leptonic sector of seesaw model.
The Yukawa terms and the mass terms for seesaw model
are given by,
\begin{eqnarray}
  {\cal L}_{m}=-y_{\nu}^{ik}\ov{L_i}N_{R_k}\tilde{\phi}
              -y_{l}^{i} \ov{L_i} l_{R_i} {\phi}
-\f{1}{2}\ov{N_{R}}^{c} M_{N_R} N_{R}+h.c., 
\label{lagrangian}
\end{eqnarray}
where $i=1,2,3$ and $k=1,2,\cdots N$. 
$L_i$ are $SU(2)$ lepton doublet fields, 
$N_{R_k}$ are the heavy Majorana right-handed
neutrinos and $l_{R_i}$ are the right-handed charged leptons.
$M_{N_R}$ is $N{\times}N$  Majorana mass matrix of
the right-handed neutrinos and is a diagonal matrix, i.e.,
$M_{N_R}=diag.(M_{1},M_{2},\cdots,M_{N})$.
$y_l$ are the Yukawa terms for charged leptons. 
We can take the basis in which both 
$M_N$ and  $y_l$ are real and diagonal without loss of generality.
In this basis,
flavor violating processes occur through off-diagonal elements of the
$3{\times}N$ Yukawa matrix
$y_{\nu}$. 
In the broken phase,
Higgs field has the vacuum expectation value $v=246$~[GeV],
and Dirac mass term is generated as
$m_{D}=\f{v}{\sqrt{2}}y_{\nu}$.

The $minimal$ seesaw model which is compatible with the present
neutrino oscillation data includes two heavy Majorana neutrinos.
For numerical analysis, we focus on the minimal model.
We summarize the results in the previous work
\cite{our-paper} which are relevant
to this paper.
There are 11 parameters in neutrino sectors 
in the model. They are two heavy Majorana masses,
$M_1, M_2$ and 9 parameters in $m_D$. 
As input parameters, we choose a heavy Majorana mass $M_1$,
the ratio of the two heavy Majorana masses, $R=\frac{M_1}{M_2}$, 
and their total decay widths $\Gamma_D^{k} (k=1,2)$.
From neutrino oscillation experiments,
two mass squared differences and
two mixing angles are obtained.
Then  8 of 11 parameters can be determined. 
The remaining three parameters are left undetermined.
They are related to
$|U_{e3}|$, CP violation of neutrino oscillation  
and neutrinoless double beta decays, which will be 
measured in the future experiments.
The formulae which relates the input parameters with $m_D$
are derived using the bi-unitary parameterization\cite{our-paper},
\begin{equation}
 m_{D}=U_{L}mV_{R},
\label{lnumass}
\end{equation}
where,
\begin{eqnarray}
U_{L}&=&\left(\begin{array}{ccc}
0 & 0 & 0 \\
0 & \cos{\theta}_{L23} & \sin{\theta}_{L23} \\
0 & -\sin{\theta}_{L23} & \cos{\theta}_{L23}
\end{array}\right)\left(\begin{array}{ccc}
\cos{\theta}_{L13} & 0 & \sin{\theta}_{L13}e^{-i{\delta}_{L}} \\
0 & 0 & 0 \\
-\sin{\theta}_{L13}e^{i{\delta}_{L}} & 0 & \cos{\theta}_{L13}
\end{array}\right) \nn \\
& &{\times}\left(\begin{array}{ccc}
\cos{\theta}_{L12} & \sin{\theta}_{L12} & 0 \\
-\sin{\theta}_{L12} & \cos{\theta}_{L12} & 0 \\
0 & 0 & 0
\end{array}\right)\left(\begin{array}{ccc}
1 & 0 & 0 \\
0 & e^{-i\f{{\gamma}_{L}}{2}} & 0 \\
0 & 0 & e^{i\f{{\gamma}_{L}}{2}} 
\end{array}\right), \label{UL}\\
m&=&\left(\begin{array}{cc}
0 & 0 \\
m_{2} & 0 \\
0 & m_{3}
\end{array}\right),
\label{mm}
V_{R}= \left(\begin{array}{cc}
\cos{\theta}_{R} & \sin{\theta}_{R} \\
-\sin{\theta}_{R} & \cos{\theta}_{R}
\end{array}\right)\left(\begin{array}{cc}
e^{-i\f{{\gamma}_{R}}{2}} & 0 \\
0 & e^{i\f{{\gamma}_{R}}{2}}
\end{array}\right).
\label{VR}
\end{eqnarray}
$m_{2}$ and $m_{3}$ are real 
and positive and $m_{2}{\leq}m_{3}$. 
Notice that there are three CP violating phases  
${\delta}_{L}, {\gamma}_{L} $ and  ${\gamma}_{R}$.

The four parameters in $m$ and $V_R$
can be written in terms of the heavy Majorana masses $M_k$, their decay widths
$\Gamma_D^k$ and the light neutrino masses $n_i (i=1,2,3)$,
\bea
\cos 2\gamma_R&=&\frac{n_2^2+n_3^2-x_1^2-x_2^2}{2(x_1x_2-n_2n_3)},\nn\\
(m_2^2,m_3^2)&=&\sqrt{M_1M_2}(\sigma_+-\rho,\sigma_++\rho),\nn\\
(\cos\theta_R, \sin\theta_R)&=&\left(\sqrt{\frac{\sigma_-+\rho}{2\rho}},
-\sqrt{\frac{-\sigma_-+\rho}{2\rho}}\right),\label{relation}
\eea
where we define the auxiliary quantities as,
\bea
x_k&=&\frac{(m_D^\dag m_D)_{kk}}{M_k}=
8\pi\Gamma_D^k\left(\frac{v}{M_k}\right)^2, \nn \\
\sigma_\pm&=&
\frac{x_2\pm x_1R}{2\sqrt{R}}, \nn \\
\rho&=&\sqrt{(x_1x_2-n_2n_3)+\sigma_-^2}.
\label{x-parameters}
\eea
The above relations follow from the eigenvalue equation
for the light neutrino mass squared $n^2$,
\bea
&& {\rm det}(m_{eff} m_{eff}^{\dagger}-n^2)=0, \nn \\
&& m_{eff} = -m_D M_{N_R}^{-1}m_D^T.
\eea
We can easily see that one of the eigenvalues ($n_1$) is zero.
It was shown that MNS matrix can be written
as a product of matrices as follows,
\bea
U_{MNS}=U_L K_R, \label{MNS}
\eea
where $K_R$ is given by,
\bea
K_R=\left(\begin{array}{ccc}
  1 &   0       & 0\\
  0 &\cos\theta &\sin\theta e^{-i\phi}\\
  0 &-\sin\theta e^{i\phi}& \cos\theta
  \end{array}  \right) 
  \left(\begin{array}{ccc}
  1 &   0       & 0\\
  0 &   e^{i\alpha}& 0\\
  0 & 0 & e^{-i\alpha}
  \end{array}  \right).
\label{kr}
\eea 
$\theta,\ \phi$ and  $\alpha$ in $K_R$ are functions of 
the parameters in $m$ and $V_R$ which are already 
determined by (\ref{relation}). Explicitly,
$K_R$ is 
obtained from the diagonalization of $m_{eff}$ as,
\bea
U_{MNS}^\dag m_{eff}U^*_{MNS}=-K_R^{\dagger}
\left(m V_RM_{N_R}^{-1}V^T_R m^T \right) K_R^*=diag.(0,n_2,n_3),
\label{DIA}
\eea
To summarize this section, we are able to determine the part of MNS
matrix, i.e., $K_R$ using the parameters $M_1$, $R$, $x_1$, $x_2$,
$n_2$ and
$n_3$. At this stage, $U_L$ is still left undetermined.
How to constrain $U_L$  will be discussed
in Section \ref{sec:nr}.

\section{Boltzmann equation}
\label{sec:be}

In this section, we shall give the derivation of 
the Boltzmann equations for lepton family asymmetries. 
The primordial lepton numbers are generated by out-of-equilibrium
decays of the heavy Majorana neutrinos. 
The decay occurs at the temperature 
higher than the electroweak phase transition
temperature ($T_{\rm EW}$), before sphaleron process being frozen.
While at extremely high temperature 
$T \ge T_{\rm sph}\simeq 10^{12}$~[GeV], the sphaleron process is not so active and the lepton number can not
be converted into the baryon number. Then, the primordial 
leptogenesis must occur at the temperature above  $T_{\rm{EW}}$ 
and below $T_{\rm sph}$ so that the baryogenesis from leptogenesis
works.
Above $T_{\rm{EW}}$,
the electroweak symmetry is recovered and the Higgs vacuum
is in the symmetric phase ($v=0$).
We focus on the Boltzmann equations which are valid
in the symmetric phase. In the phase, the lepton family number 
violating processes involve the Yukawa coupling of the neutrinos
$y_{\nu}$.
Using the Boltzmann equation,
we can trace the evolution of the lepton family number asymmetries
from the mass scales of heavy Majorana neutrinos down to $T_{\rm{EW}}$.
In order to trace the evolution into the lower temperature regime
$T \le T_{\rm{EW}}$,
the Boltzmann equations in the broken phase must be used. 
The derivation of the equations in the broken phase 
is beyond the scope of the present paper.

We first show the Boltzmann equations for the heavy Majorana neutrino
number densities $n_k$ ($k=1,2,\cdots N$) and lepton family 
number  densities $n_{L_i}$ ($i=1,2,3$),
\bea
\frac{dn_k}{dt}+3 H(z) n_k&=&C_{n_k},\nn\\
\frac{dn_{L_i}}{dt}+3 H(z) n_{L_i} &=&C_{l_i},
\label{general}
\eea
where $M_1$ is the mass of the lightest heavy Majorana neutrino and 
$z=\frac{M_1}{T}$.  $H(z)=H(1)z^{-2}$ is the Hubble parameter with 
$H(1)=\sqrt{\frac{4 \pi^3 g_*}{45}}\frac{M_1^2}{m_P}$. 
$m_P$ denotes the Planck mass scale
and $g_*\simeq 106.75$ counts the total number of light particles
degrees of freedom at $T_{\rm EW} \le T \le T_{\rm sph}$.
The right-hand side of the Boltzmann equations
$C_{n_k}$ and $C_{l_i}$ are source terms
and they denote the particle number changes per unit spacetime volume.
The source terms are obtained by computing the decays, the inverse decays,
and the scattering processes.

In deriving the Boltzmann equations, we use Maxwell-Boltzmann distributions
for light particles. Explicitly, the phase 
space density of a
light particle $X$  with energy $E_X$ and chemical potential $\mu_X$
is given by,
\bea
f_X={\rm exp}\left[-\frac{E_X-\mu_X}{T}\right].
\label{fxsm}
\eea
The distribution is a good approximation in the absence of Bose condensation 
and Fermi degeneracy.
The signs of the 
chemical potentials for particle and anti-particle are opposite
each other. The up component in $SU(2)$ doublets 
has the same chemical potential as that of the down component
in the symmetric phase of electroweak
symmetry. For the heavy Majorana neutrino $N_{R_k}$, we assume
that the distribution is proportional to the equilibrium distribution.
Under this assumption, the normalization constant
is given by the ratio of the non-equilibrium number density $n_k$ 
and the equilibrium density $n_k^{eq}$,
\bea
\label{fn}
f_{N}(s_z=\frac{1}{2})=f_{N}(s_z=-\frac{1}{2})\equiv\frac{n_k}{n_k^{eq}}e^{-E_N/T},\\
n^{eq}_k=2\int\frac{d^3 p_N}{(2 \pi)^3} e^{-E_N/T}=\frac{1}{\pi^2}
\frac{M_{k}^3}{z \sqrt{a_k}}K_2(z\sqrt{a_k}),
\label{nkeq}
\eea
where  $s_z$ denotes  the spin of the heavy Majorana neutrinos.
$a_k=M_{k}^2/M_{1}^2$. $K_1$ and $K_2$ are modified Bessel functions.

\subsection{Calculation of $C_{n_k}$ and $C_{l_i}$} 
\label{sub1}

The processes contributing to the source terms
can be divided into two
parts. One includes the decays and inverse decays of the heavy Majorana
neutrinos, and the other includes 
two-body scattering. The contribution from decay and inverse decays
are denoted as  $C^D$
and the contribution from scattering processes are
denoted as $C^R$.
Then the total source terms are given by their sum,
$C_{n_k}=C_{n_k}^{D}+C_{n_k}^{R}$  for the heavy Majorana 
neutrinos densities and 
$C_{l_i}=C_{l_{i}}^{D}+C_{l_{i}}^{R}$ 
for the lepton family number densities respectively.

Below, in the results of the two-body scattering $a+b\to c+d$,
we define the reaction rate
$\gamma$ as, 
\bea
\gamma=\frac{M_{1}^4}{64\pi^4}\frac{1}{z}
\int\limits_{(m_a+m_b)^2/M_{1}^2}^{\infty} 
d\hat{s}\hat{\sigma}(\hat{s}) \sqrt{\hat{s}}K_1(z\sqrt{\hat{s}}),
\label{resc}
\eea 
where $\hat{\sigma}(\hat{s})$ is the reduced cross section and 
$\hat{s}=(p_{a}+p_{b})^{2}/M_{1}^2$.
The reduced cross section $\hat{\sigma}(\hat{s})$ 
is related to the 
usual cross section ${\sigma}(\hat{s})$ by
$\hat{\sigma}(\hat{s})=\frac{8}{(p_a+p_b)^2}
[(p_a\cdot p_b)^2-m_a^2m_b^2]\sigma(\hat{s})$.
All the reduced cross sections $\hat{\sigma}$ are 
collected in Appendix B.

%%%%%%%%%
\begin{wraptable}{r}{\halftext}
\caption{
The decays and inverse decays of the heavy neutrinos contributing to 
lepton and heavy Majorana neutrino number densities.}
\label{Tab1} 
\begin{center}
\begin{tabular}{|c|c|c|}\hline\hline
\m{Number change} &  \m{Processes} & \\ \hline
${\Delta}L=1,~{\Delta}N=1 $&  $N_{R_{k}}~{\leftrightarrow}~l_i^{\pm}{\phi}^{\mp}$ & \\ 
 &  $N_{R_{k}}~{\leftrightarrow}~{\nu}_{i}{\phi}^{0}$ &
 ${\gamma}_{D}^{ki}$ \\ 
 &  $N_{R_{k}}~{\leftrightarrow}~\ov{\nu}_{i}{{\phi}^{0}}^{*}$ & \\ \hline
source term & $C_{n_{k}}^{D},~C_{l_{i}}^{D}$ & \\ 
\hline\hline
\end{tabular}
\end{center}
\end{wraptable}

\begin{figure}[htb]
\begin{center}
\epsfig{file=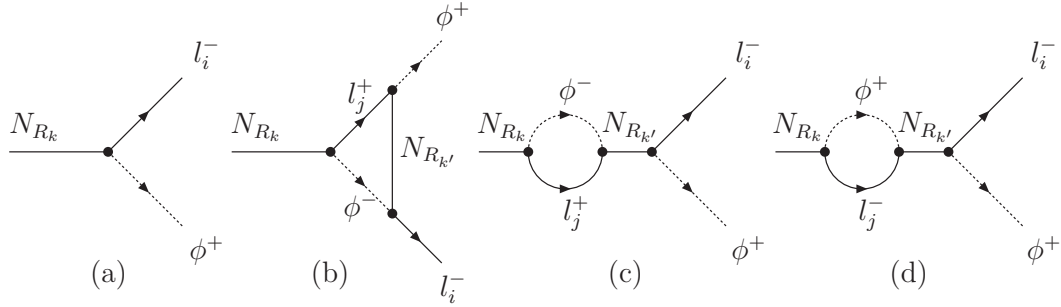,width=14cm}
\caption{Feynman diagrams of the heavy Majorana neutrino decay: (a)
Tree level diagram.   One loop diagrams: (b) vertex type and (c) (d)
self-energy type diagrams.}
\label{fey1}
\end{center}
\end{figure}

Firstly, we calculate the source term $C_{n_k}$ in 
(\ref{general}).
The decays and the inverse decays of the heavy Majorana neutrinos 
contribute to the source term. 
The processes involved in the calculation are listed in Table \ref{Tab1}.
Using the Lagrangian (\ref{lagrangian}), the definitions in Appendix A,
and the phase space densities in 
(\ref{fxsm}) and (\ref{fn}),  we calculate the Feynman diagrams shown 
in Fig.~\ref{fey1}(a) and obtain,
\bea
C_{n_k}^D&=&-\sum\limits_i\gamma_D^{ki}
\left(\frac{n_k}{n_k^{eq}}-\cosh\frac{\mu_{\phi^+}+\mu_{l^-_{iL}}}{T}
\right),
\eea
where $\gamma^{ki}_D$ is related to the partial decay width
$\Gamma_{ki}(N_{R_k}\to l_i^-\phi^+)$,
\bea 
\gamma^{ki}_D=4n_k^{eq}
\Gamma_{ki}\frac{K_1(z\sqrt{a_k})}{K_2(z\sqrt{a_k})}, \ \ 
\label{gammadki}
\Gamma_{ki}=\f{1}{32{\pi}}\left|(y_{\nu})_{ik}\right|^{2}M_{k}.
\eea
\begin{wrapfigure}{r}{6.6cm}
\centerline{\includegraphics[width=6.0cm]{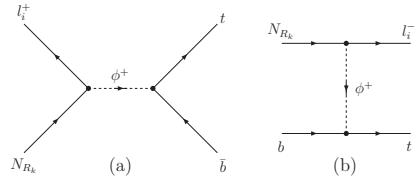}}
\caption{ Processes contributing
 to both the heavy Majorana neutrino densities and lepton number 
asymmetries with Higgs exchanges via (a) s channel and (b) t channel.}
\label{fey3}
\end{wrapfigure}

\begin{table}[htb]
\caption{\label{Tab3} 
The processes with Higgs exchange contributing to the lepton family
 number densities and heavy Majorana neutrino number densities.}
\begin{center}
\begin{tabular}{|c|c|c|c|c|}\hline\hline
Number change&Processes via t channel & & Processes via s channel & \\ \hline
${\Delta}L=1,~{\Delta}N=1$ &
 $N_{R_{k}}\ov{t}_{R}~{\leftrightarrow}~l_{i}^{-}\ov{b}_{L}$ & &$ N_{R_{k}}l_{i}^{+}~{\leftrightarrow}~\ov{b}_{L}t_{R}$ &  \\
 &$N_{R_{k}}b_{L}~{\leftrightarrow}~l_{i}^{-}t_{R}$& &
$ b_{L}\ov{t}_{R}~{\leftrightarrow}~N_{R_{k}}l_{i}^{-}$ & \\
 &$l_{i}^{+}b_{L}~{\leftrightarrow}~N_{R_{k}}t_{R}$& &
$t_{L}\ov{t}_{R}~{\leftrightarrow}~N_{R_{k}}{\nu}_{i}$ & \\
 & $l_{i}^{+}\ov{t}_{R}~{\leftrightarrow}~N_{R_{k}}\ov{b}_{L}$& ${\gamma}_{{\phi},t}^{ki}$ &
$N_{R_{k}}\ov{\nu}_{i}~{\leftrightarrow}~\ov{t}_{L}t_{R} $ & ${\gamma}_{{\phi},s}^{ki}$\\
 &$ N_{R_{k}}t_{L}~{\leftrightarrow}~t_{R}{\nu}_{i}$& & & \\
 &$ \ov{t}_R\ov{\nu}_{i}~{\leftrightarrow}~N_{R_{k}}\ov{t}_{L}$& & &\\
 &$N_{R_{k}}\ov{t}_{R}~{\leftrightarrow}~\ov{t}_{L}{\nu}_{i}$& & & \\
 &$t_{L}\ov{\nu}_{i}~{\leftrightarrow}~N_{R_{k}}t_{R}$& & & \\ \hline
source term & $C_{nk}^{(1)R},~C_{l_{i}}^{(1)R}$ &  &
 $C_{nk}^{(2)R},~C_{l_{i}}^{(2)R}$ & \\  

\hline\hline
\end{tabular}
\end{center}
\end{table}

Since the Yukawa coupling of top quark 
is much larger than those of the other SM particles, 
the heavy Majorana neutrino creation and annihilation 
processes in Fig. \ref{fey3}
give important contributions to the heavy 
Majorana neutrino number densities. \cite{Plumacher}
The processes are listed in Table \ref{Tab3}.
Their contributions to the source term $C_{n_k}$ are given by,
\bea
C_{n_k}^R&=&C_{n_k}^{(1)R}+C_{n_k}^{(2)R},
\eea
where $C_{n_k}^{(1)R}$ and $C_{n_k}^{(2)R}$ are,
\bea
C_{n_k}^{(1)R}&=&-\sum\limits_i \gamma^{ki}_{\phi,t}
\left[\frac{n_k}{n_k^{eq}}\left(\cosh\frac{\mu_{t_L}}{T}
+\cosh\frac{\mu_{t_R}}{T}\right)\right.\nn\\
&&\left.-\cosh\frac{\mu_{l^-_{iL}}+\mu_{t_L}}{T}
-\cosh\frac{\mu_{l^-_{iL}}-\mu_{t_R}}{T}\right],\\
C_{n_k}^{(2)R}&=&-\sum\limits_i \gamma^{ki}_{\phi,s}
\left[\frac{n_k}{n_k^{eq}}
-\cosh\frac{\mu_{\phi^+}+\mu_{l_{iL}^-}}{T}\right].
\eea
The reaction rates  $\gamma_{\phi,s}^{ki}$ and $\gamma^{ki}_{\phi,t}$ 
are related
to  the reduced cross sections $\hat{\sigma}^{ki}_{\phi,s}$ and
$\hat{\sigma}^{ki}_{\phi,t}$.  Their explicit form are given in
(\ref{sigma-phis}) and (\ref{sigma-phit}). The corresponding processes
are shown in Fig. \ref{fey3}(a) and Fig. \ref{fey3}(b), respectively. 

Next let us study the source terms for lepton family numbers, $C_{l_i}$ in (\ref{general}).  
The decays and the inverse decays contribution to $C_{l_{i}}^D$ is
given by,
\bea
C_{l_{i}}^{D}&=&\sum\limits_k\gamma_D^{ki}\left[\epsilon_i^k
\left(\frac{n_k}{n_k^{eq}}+\cosh\frac{\mu_{\phi^+}+\mu_{l^-_{iL}}}
{T}\right)+\sinh\frac{\mu_{\phi^-}+\mu_{l^+_{iL}}}{T}\right],
\label{cld}
\eea
where the lepton family CP  asymmetry $\epsilon_i^k$  is defined as,  
\bea
\epsilon_i^k=
\frac{\Gamma(N_{R_k}\to l^-_i\phi^+)-\Gamma(N_{R_k}\to l^+_i\phi^-)}
{\Gamma(N_{R_k}\to l^-_i\phi^+)+\Gamma(N_{R_k}\to l^+_i\phi^-)}.
\eea
The CP asymmetries are generated by the interference of the tree diagram 
and one-loop diagrams shown in Fig. \ref{fey1}, 
\bea
\epsilon_i^k&=&\frac{1}{8\pi}
\sum\limits_{k'\neq k}\left[I(x_{k'k})
\frac{\Im[(y_\nu^\dag y_\nu)_{kk'}(y_\nu)^*_{ik}(y_\nu)_{ik'}]}
{|(y_\nu)_{ik}|^2}\right.\nn\\ 
&&\left.+\frac{1}{1-x_{k'k}}
\frac{\Im[(y_\nu^\dag y_\nu)_{k'k}(y_\nu)^*_{ik}(y_\nu)_
{ik'}]}{|(y_\nu)_{ik}|^2}\right],
\label{epsilon}
\eea
where $x_{k'k}=M_{k'}^2/M^2_{k}$ and
\bea
I(x)&=&\sqrt{x}\left[1+\frac{1}{1-x}+(1+x)\ln\frac{x}{1+x}\right]\nn\\
     &=&\left\{\begin{array}{ll}
        -\frac{3}{2}x^{-1/2}& \ \ \ for \ \ x\gg 1,\\
        -2x^{3/2}           & \ \ \ for \ \ x\ll 1.  
              \end{array}\right.
\label{IX}
\eea
The diagram Fig.~\ref{fey1}(d)
gives the contributions to the second terms of the
lepton family CP asymmetry (\ref{epsilon}). The diagram
does not contribute to the total lepton asymmetry while
it does contribute to the lepton family CP asymmetries.

The contributions to the source term $C_{l_i}$ 
from two particle scatterings can be divided into four 
parts,
\begin{equation}
 C_{l_{i}}^{R}=\sum\limits_{\chi=1}^4 C_{l_{i}}^{(\chi)R}.
\label{315}
\end{equation}
The first two terms in (\ref{315}) are obtained by calculating the diagrams
in Fig. 2. 
They are given as follows,
\bea   
C_{l_{i}}^{(1)R}\left[\mbox{Fig. \ref{fey3}(b)} \right]&=&\sum\limits_k\gamma^{ki}_{\phi,t}
\left[\frac{n_k}{n_k^{eq}}\left(\sinh\frac{\mu_{t_L}}{T}
 -\sinh\frac{\mu_{t_R}}{T}\right) \right. \nn \\
&& \left.+\sinh\frac{\mu_{l^+_{iL}}+\mu_{t_L}}{T}
+\sinh\frac{\mu_{l^+_{iL}}-\mu_{t_R}}{T}\right],\\
C_{l_{i}}^{(2)R}\left[\mbox{Fig. 2(a)} \right]&=&\sum\limits_k\gamma^{ki}_{\phi,s}
\left[\frac{n_k}{n_k^{eq}}\sinh\frac{\mu_{l^+_{iL}}}{T}
+\sinh\frac{\mu_{t_L}-\mu_{t_R}}{T}\right].
\label{cr12}
\eea
The processes listed in Table III also contribute to the
$C_{l_{i}}^{R}$. 
They are denoted by $C_{l_{i}}^{(3)R}$ and
$C_{l_{i}}^{(4)R}$ and are given by,
%%%%%%
%%%%%%
\begin{table}[htb]
\begin{center}
\caption{$|\Delta L|=2$ and $\Delta L=0$ lepton family number changing
scattering processes}
\begin{tabular}{|c|c|c|c|c|}\hline\hline
Number change & Processes via s (u) channel & & Process via t (u)
 channel & \\  \hline
$|{\Delta}L=2|,\ {\Delta}N=0$ &
 $~~l_{i}^{\pm}{\phi}^{\mp}~{\leftrightarrow}~l_{j}^{\mp}{\phi}^{\pm}$ & 
 & ${\phi}^{0*}{\phi}^{0*}~{\leftrightarrow}~{\nu}_{i}{\nu}_{j}$ & \\ 
 & $\ov{\nu}_{j}{\phi}^{0*}~{\leftrightarrow}~{\nu}_{i}{\phi}^{0}$ & 
 ${\gamma}_{N,1}^{ij}$&
 $l_{i}^{\pm}l_{j}^{\pm}~{\leftrightarrow}~{\phi}^{\pm}{\phi}^{\pm}$ &
 ${\gamma}_{N,2}^{ij}$ \\ 
 & $\ov{\nu}_{i}{\phi}^{0*}~{\leftrightarrow}~{\nu}_{j}{\phi}^{0}~~$ &
 & ${\phi}^{0}{\phi}^{0}~{\leftrightarrow}~\ov{\nu}_{i}\ov{\nu}_{j}$ &
 \\ \cline{2-5}
& $\ov{\nu}_{i}{\phi}^{0*}~{\leftrightarrow}~l_{j}^{-}{\phi}^{+}$ & & 
 $l_{i}^{-}{\nu}_{j}~{\leftrightarrow}~{\phi}^{-}{\phi}^{0*}$ &
 \\
 & $l_{j}^{+}{\phi}^{-}~{\leftrightarrow}~{\nu}_{i}{\phi}^{0}$ & $
 \gamma_{N,3}^{ij}$&
 $l_{i}\ov{\nu}_{j}~{\leftrightarrow}~{\phi}^{+}{\phi}^{0}$ & ${\gamma}_{N,t1}^{ij}$\\
&$l_{i}^{+}{\phi}^{-}~{\leftrightarrow}~{\nu}_{j}{\phi}^{0}$  & & 
 $l_{j}^{-}{\nu}_{i}~{\leftrightarrow}~{\phi}^{-}{\phi}^{0*}$ & \\
& $\ov{\nu}_{j}{\phi}^{0*}~{\leftrightarrow}~l_{i}^{-}{\phi}^{+}$ & & 
 $l_{j}^{+}\ov{\nu}_{i}~{\leftrightarrow}~{\phi}^{+}{\phi}^{0}$ & \\ 
 \cline{1-5} $\Delta L=0,\ \Delta L_i \ne 0$
& $l_{i}^{\pm}{\phi}^{\mp}~{\leftrightarrow}~l_{j}^{\pm}{\phi}^{\mp}$ & &
 ${\phi}^{+}{\phi}^{-}~{\leftrightarrow}~l_{i}^{\pm}l_{j}^{\mp}$ &
 \\
& $\ov{\nu}_{j}{\phi}^{0*}~{\leftrightarrow}~l_{i}^{+}{\phi}^{-}$ & &
 ${\nu}_{i}\ov{\nu}_{j}~{\leftrightarrow}~{\phi}^{0}{\phi}^{0*}$ & \\
& $\ov{\nu}_{i}{\phi}^{0*}~{\leftrightarrow}~\ov{\nu}_{j}{\phi}^{0*}$ & &
 ${\nu}_{j}\ov{\nu}_{i}~{\leftrightarrow}~{\phi}^{0}{\phi}^{0*}$ & \\
 & $\ov{\nu}_{j}{\phi}^{0}~{\leftrightarrow}~l_{i}^{-}{\phi}^{+}$ &${\gamma}_{N,s}^{ij}$ &
$\nu_i l_j^+~{\leftrightarrow}~\phi^+ \phi^{0*}$ & ${\gamma}_{N,t2}^{ij}$\\
& $\ov{\nu}_{i}{\phi}^{0*}~{\leftrightarrow}~l_{j}^{-}{\phi}^{+}$ 
& & $\ov{\nu_i} l_j^-~{\leftrightarrow}~\phi^0 \phi^-$
& \\
& ${\nu}_{j}{\phi}^{0}~{\leftrightarrow}~{\nu}_{i}{\phi}^{0}$ &
& $ \ov{\nu_j} l_i^-~{\leftrightarrow}~\phi^{0}\phi^-$  & \\
& $l_{j}^{-}{\phi}^{+}~{\leftrightarrow}~{\nu}_{i}{\phi}^{0}$ &&
$ \nu_j l_i^+{\leftrightarrow}~\phi^+ \phi^{0*} $ & \\ \hline
source term & $C_{l_{i}}^{(3)R}$ & &  $C_{l_{i}}^{(4)R}$ & \\ 
\hline\hline 
\end{tabular}
\end{center}
\label{Tab2} 
\end{table}

\bea
C_{l_{i}}^{(3)R}&=&-\sum\limits_{j,k}
\frac{|(\ynu)_{jk}|^2}{(\ynu^\dag\ynu)_{kk}}\gamma_D^{ki}
\left[\epsilon_i^k\left(\cosh\frac{\mu_{\phi^-}+\mu_{l_{iL}^+}}{T}+
\cosh\frac{\mu_{\phi^-}+\mu_{l_{jL}^+}}{T}\right)
\right.\nn\\
&&\left.+\sinh\frac{\mu_{\phi^-}+\mu_{l_{iL}^+}}{T}\right]
+\sum\limits_{j}(\gamma^{ij}_{N,1}+\gamma^{ij}_{N,3})\left(
\sinh\frac{\mu_{\phi^-}+\mu_{l_{iL}^+}}{T}+
\sinh\frac{\mu_{\phi^-}+\mu_{l_{jL}^+}}{T}\right)\nn\\
&&+2\sum\limits_{j}\gamma^{ij}_{N,s}\left(
\sinh\frac{\mu_{\phi^-}+\mu_{l_{iL}^+}}{T}-
\sinh\frac{\mu_{\phi^-}+\mu_{l_{jL}^+}}{T}\right),\label{crl1} \\
C_{l_{i}}^{(4)R}&=&
\sum\limits_{j}\left[\left(\frac{1}{2}\gamma^{ij}_{N,2}
+\gamma^{ij}_{N,t1}\right)
\left(
\sinh\frac{2\mu_{\phi^-}}{T}+
\sinh\frac{\mu_{l^+_{iL}}+\mu_{l_{jL}^+}}{T}\right)
\right.\nn\\
&&\left.+2 \gamma^{ij}_{N,t2}
\sinh\frac{\mu_{l_{iL}^+}-\mu_{l_{jL}^+}}{T}\right].
\label{crl2}
\eea
The processes in Table III are classified into the total lepton number
changing scatterings ($|\Delta L|=2$) and 
the total lepton number conserving scatterings ($\Delta L=0$). 
\begin{wrapfigure}{r}{6.6cm}
\centerline{\includegraphics[width=6cm]{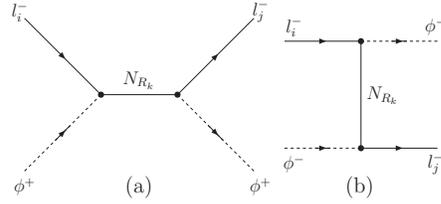}}
\caption{The lepton family number violating scatterings 
with  $N_{R_k}$ exchanges via (a) s channel and (b) u channel.}
\label{fey2}
\end{wrapfigure}
In the former, the lepton family number $L_i$
changes by one unit or by two units.
The latter corresponds to $|\Delta L_i|=1$.
The total lepton number conserving and the 
lepton family number violating processes
must be taken into account when
we study each lepton family number evolution.
In Fig. 3, the typical Feynman diagrams which
correspond to $\Delta L=0$
$\Delta L_i=-1$ are shown.
Their reaction rates are included in $\gamma_{N,s}$
and $\gamma_{N,t2}$ of (\ref{crl1}) and (\ref{crl2}).
Finally we also note that in (\ref{crl1}), 
the term ${\epsilon}_i^k$ comes from the on-shell
contribution of 
s-channel heavy Majorana particle exchanged diagrams.
%%%%%%%%%%%%%%%%%%%%%%%%%%%%%%%%%%%%%%%%%%%%%%%%%%%%%
%%%%%%%%%%%%%%%%% <Sphaleron> %%%%%%%%%%%%%%%%%%%%%%%
%%%%%%%%%%%%%%%%%%%%%%%%%%%%%%%%%%%%%%%%%%%%%%%%%%%%%
\subsection{Sphaleron and chemical potential relations}
\label{sec:sph}

In the previous subsection, we write the source terms with
the reaction rates $\gamma$ and the chemical potentials $\mu$. 
In this subsection,
we derive the relation between the lepton family asymmetries $Y_{Li}$ and
chemical potentials by considering the various
equilibrium conditions. We adapt the method developed by
Harvey and Turner. 
~\cite{HarveyTurner}~\cite{bento}

At first, let us consider the effective action of sphaleron for all
left-handed fermions,
\begin{equation}
 {\cal O}_{B+L}=\prod_{i}(Q_{i}Q_{i}Q_{i}L_{i}),
\end{equation} 
where $Q_{i}, L_{i}$ are quark and lepton doublets respectively.
The sphaleron transition rate at high temperature, in the symmetric phase, 
is estimated to be~\cite{sphaleron-rate}
\begin{equation}
 {\Gamma}_{\rm sph}~{\sim}~2{\times}10^{2}{\alpha}_{W}^{5}T,
\end{equation}
where ${\alpha}_{W}$ is weak coupling constant at Grand Unified scale
and is taken to be $\frac{1}{40}$.
From the thermal equilibrium condition, i.e., ${\Gamma}_{\rm sph}>H$
, we obtain the upper bound of the temperature
$T_{\rm sph}=1.4{\times}10^{12}~{\rm [GeV]}$. Above the temperature,
sphaleron process is not active. At low energy, below $T_{\rm EW}
\simeq 10^2~[\rm{GeV}]$,
the sphaleron process is frozen out again, then the active
temperature range of sphaleron is, 
\begin{equation}
T_{\rm EW} \le T \le T_{\rm sph}.
\label{sphbound}
\end{equation}
In the temperature range (\ref{sphbound}),
the sphaleron equilibrium condition leads to the chemical potential relation,
\begin{equation}
\sum_{i=1}^{N_f}
\left( {\mu}_{u_{iL}}+2{\mu}_{d_{iL}}+{\mu}_{i{\nu}}\right)=0.
\label{spha-con}
\end{equation}
where $N_f=3$.
We assume that
gauge and Yukawa interactions are in equilibrium. 
From charged current interaction processes, we obtain the
following chemical potential relations,
\begin{equation}
 {\mu}_{W^{-}}={\mu}_{d_{iL}}-{\mu}_{u_{iL}}={\mu}_{l^{-}_{iL}}-{\mu}_{{\nu}_{iL}}={\mu}_{{\phi}^{-}}+{\mu}_{{\phi}^{0}},
\label{chemicalW}
\end{equation}
where $i~(i=1,2,3)$ denote indices for the generations. 
By taking the weak basis that the Yukawa term for up type quarks is
flavor diagonal and that for down type quarks has
flavor off-diagonal elements,
we obtain the chemical potential relations 
from the equilibrium condition of the Yukawa interactions,
\begin{equation}
 {\mu}_{{\phi}^{0}}= {\mu}_{u_{iR}}-{\mu}_{u_{iL}}={\mu}_{d_{iL}}-
 {\mu}_{d_{jR}}={\mu}_{l^{-}_{iL}}-{\mu}_{l^{-}_{iR}},
\label{chemicalY}
\end{equation}
where we note that the
chemical potentials for the left-handed down quarks ($d_{iL}$)
and those of 
the right-handed down quarks ($d_{jR}$) satisfy the
flavor mixed relations.
Therefore the chemical potentials
for the right-handed down quarks and the left-handed down quarks
are flavor independent. Both of them
can be written by a single chemical potential as,
\bea
\mu_{d_{iR}}=\mu_{d_R}, \quad
\mu_{d_{iL}}=\mu_{d_L}. \label{chemicald}
\eea 
Using (\ref{chemicalW}),(\ref{chemicalY}) and (\ref{chemicald}), we
can also show that the chemical potentials for up type quarks
are flavor independent,
\bea
\mu_{u_{iR}}=\mu_{u_R}, \quad 
\mu_{u_{iL}}=\mu_{u_L}. 
\label{chemicalup}
\eea
Next, we can
relate the chemical potentials of SU(2) doublets.
This follows from the chemical potential 
for $W$ boson vanishes in the symmetric phase.~\cite{HarveyTurner}. 
From (\ref{chemicalW}), the up and down components of 
SU(2) doublets have the same chemical potential. 
Note that the chemical potentials for 
leptons can be flavor dependent.
\bea
{\mu}_{l^{-}_{iL}}={\mu}_{{\nu}_{iL}}={\mu}_{l^{-}_{iR}}+\mu_{\phi^0}.
\label{W-boson=0}
\eea
We also take into account of the charge neutrality condition. 
The condition can be written as,
\bea
Q&=& \f{T^{2}}{3}\left[\sum_{i=1}^{N_{f}}{\mu}_{u_{iL}}-{\mu}-\left(2N_{f}+N_{H}+2\right){\mu}_{W^{-}}+(2N_f+N_{H}){\mu}_{{\phi}^{0}}\right]=0,\nn\\ 
\label{charge}
\eea
where ${\mu}{\equiv}\sum\limits_{i}{\mu}_{{\nu}_{iL}}$ and 
$N_H$ is the number of Higgs doublet and is taken to be $1$.
With (\ref{spha-con}),(\ref{chemicald}),(\ref{chemicalup}),
(\ref{charge}) and $\mu_W=0$, we obtain the following relations,
\bea
&& 9 \mu_{\mu_L} + \mu =0 \nn \\
&& 3 \mu_{\mu_L}=\mu-7 \mu_{\phi^0}.
\eea
Then the chemical potentials for quarks and charged leptons
can be written by the single chemical potential 
of $\mu=\sum\limits_{i}{\mu}_{{\nu}_{iL}}$ of neutrinos as follows:
\bea
\begin{array}{lll}
\mu_{u_L}=\mu_{d_L}=-\frac{1}{9}\mu,&
\mu_{u_R}=\frac{5}{63}\mu,&\mu_{d_R}=-\frac{19}{63}\mu,\\
\mu_{\phi^0}=-\mu_{\phi^-}=\frac{4}{21}\mu,& 
\sum\limits_i\mu_{l^-_{iL}}=\mu,\\
\sum\limits_i\mu_{l^-_{iR}}=\frac{3}{7}\mu.\\
\end{array}
\label{chemical1}
\eea
We can also write the baryon number density and lepton number
densities with the chemical potentials as,
%The number density of baryon $n_{B}$, lepton $n_{L}$, 
%${\mu}_{W}$, ${\mu}_{{\phi}^{0}}$, ${\mu}_{u_{L}}$ and ${\mu}_{{\nu}_{iL}}$,
\begin{eqnarray}
n_{Li}&=&\f{T^{2}}{6}(\mu_{{\nu}_{iL}}+{\mu}_{l_{iL}^{-}}+{\mu}_{l_{iR}^{-}})=\f{T^{2}}{2}\left({\mu}_{l_{iL}^{-}}-\f{4}{63}{\mu}\right), \nn \\
n_{L}&=& \sum_{i=1}^{N_{f}}n_{Li}=\f{T^{2}}{2}\left({\mu}-\f{4}{21}{\mu}\right)=\f{17}{42}T^{2}{\mu}, \\
n_{B}&=& \f{T^{2}}{6}\sum_{i=1}^{N_{f}}({\mu}_{u_{iL}}+{\mu}_{u_{iR}}+{\mu}_{d_{iL}}+{\mu}_{d_{iR}})=N_f\left(\f{2}{3}T^{2}{\mu}_{u_{L}}\right)=-\f{2}{9}T^{2}{\mu}. \nn
\label{chemical}
\end{eqnarray}

By using the equations above, we finally obtain the relations between
the chemical potentials for the lepton doublets ($\mu_{l^-_{iL}}$)
and lepton family asymmetries,
\begin{eqnarray}
Y_{L_{i}}&=&\f{n_{L_{i}}}{s}=\f{1}{2}\left(\f{{\mu}_{l_{iL}^{-}}}{T}\right)\f{T^{3}}{s}-\f{4}{51}Y_{L}, \nn \\
Y_{L}&=&\f{n_{L}}{s}=\f{17}{42}\left(\f{{\mu}}{T}\right)\f{T^{3}}{s},
\label{chemical2}
\end{eqnarray}
where $s$ is entropy density given by
$s=\frac{2\pi^2}{45}g_*\left(\frac{M_1}{z}\right)^3.$ 
What is important particularly from the flavor-dependent point of view
is that family asymmetry $Y_{L_{i}}$ take a different value on the
each generation.
The  conversion rate from lepton asymmetry to baryon asymmetry
is given by the well known formulae \cite{HarveyTurner},
\bea
Y_{B}=-\f{28}{51}Y_{L}. \label{conversion}
\eea
We will use the relations 
(\ref{chemical1}) and (\ref{chemical2}) to express all the chemical
potentials in terms of the lepton family asymmetries in the next subsection.
\subsection{Boltzmann Equations}
Now we can write the Boltzmann equations in a tractable form.
Using the definition in (\ref{chemical}),$Y_{N_k}=\frac{n_k}{s}$ and
$Y_{L}^{eq}=\f{T^{3}}{s}$,
the Boltzmann equations (\ref{general}) can be rewritten as, 
\begin{eqnarray}
\f{dY_{N_{k}}}{dz}&=& -\f{z}{sH(1)}\left(\f{Y_{N_{k}}}{Y_{N_{k}}^{eq}}-1\right)\sum_{i}\left[{\gamma}_{D}^{ki}+{\gamma}_{{\phi},s}^{ki}+2{\gamma}_{{\phi},t}^{ki}\right], \nn \\
\f{dY_{L_{i}}}{dz}&=& -\f{z}{sH(1)}\left\{-\sum_{k}\left(\f{Y_{N_{k}}}{Y_{N_{k}}^{eq}}-1\right){\epsilon}_{i}^{k}{\gamma}_{D}^{ki} \right. \nn \\ 
&& \left. +\sum_{j}\f{Y_{L_{j}}}{Y_{L}^{eq}}\left[A_{ij}+\sum_{k}\f{Y_{N_{k}}}{Y_{N_{k}^{eq}}}A_{ij}^{k}\right]\right\},
\label{ynkyli}
\end{eqnarray}
where the time $t$ derivative in (\ref{general})
is replaced by derivative on $z$ using
the relation in the radiation-dominated epoch,
\bea
t&=&\frac{1}{2 H(1)} z^2, \nn \\ 
H(1)&=&\sqrt{\frac{4 \pi^3 g_*}{45}}\frac{M_1^2}{m_P}.
\eea
In the source terms $C_{l_i}^{D,R}$, $C_{n_k}^{D,R}$ derived in
subsection \ref{sub1}, we adapt the chemical potential relations in
(\ref{chemical1}) and (\ref{chemical2}),
and the approximation of  $\sinh\frac{\mu_X}{T}\simeq\frac{\mu_X}{T}$, 
\  $\cosh\frac{\mu_X}{T}\simeq 1$. Then
 $A_{ij}$ and $\ A_{ij}^{k}$ are given as follows,
\begin{eqnarray}
A_{ij}&=& 16\sum_{j'}^{3}
\left[\left(\f{1}{8}({\delta}_{ij}+{\delta}_{jj'})+\f{4}{51}\right)
\left({\gamma}_{N,1}^{ij'}+{\gamma}_{N,3}^{ij'}\right)
+\left(\f{2}{51}
+\f{1}{16}({\delta}_{ij}+{\delta}_{jj'})
\right){\gamma}_{N,2}^{ij'}\right. \nn \\
&& \left.
+\left(\f{1}{8}({\delta}_{ij}+\delta_{jj'})+\f{4}{51}\right){\gamma}_{N,t1}^{ij'}
+\f{1}{4}({\delta}_{ij}-\delta_{jj'})\left({\gamma}_{N,t2}^{ij'}
+{\gamma}_{N,s}^{ij'}\right)\right] \nn \\
&&
+8\sum_{k}\left[\left(\f{1}{2}{\delta}_{ij}+\f{5}{51}\right){\gamma}_{{\phi},t}^{ki}+\f{1}{17}{\gamma}_{{\phi},s}^{ki}\right], \label{aij} \\
A_{ij}^{k}&=&2\left[\f{4}{17}{\gamma}_{{\phi},t}^{ki}+\left({\delta}_{ij}+\f{4}{51}\right){\gamma}_{{\phi},s}^{ki}\right].
\label{aijk}
\end{eqnarray}
%If only left-handed  lepton chemical potentials are taken to non-zero, 
%after summing  all the generations for leptons, our formulae are in 
%agreement  with those of Ref.\ \cite{Plumacher}. Note there exist 
%some discrepancies in expressions of the reduce cross sections 
%except for the process $\Ntlmb$ presented in Appendix B.

%%%%%%%%%%%%%%%%%%%%%%%%%%%%%%%%%%%%%%
%%  Numerical results %%%%%%%%%%%%%%%%%%
%%%%%%%%%%%%%%%%%%%%%%%%%%%%%%%%%%%%%%%
\section{Numerical results}\label{sec:nr}
In this section, we present the numerical results for the
lepton family number
asymmetries based on the minimal seesaw model described in Sec. 
\ref{sec:model}. As we showed in the previous section, the Yukawa coupling
$y_{\nu}=\frac{\sqrt{2}m_D}{v}$,
and the heavy Majorana masses $M_1$,$M_2$ are needed for 
the numerical analysis
of Boltzmann equation. We parameterize $m_D$ 
in the bi-unitary form, i.e., $m_D=U_L m V_R$ and write the parameters
in $m$ and $V_R$ with $M_1,R=\frac{M_1}{M_2},
x_1,x_2,n_2$ and $n_3$.

 Next we show how the angles in $U_L$
are constrained by discussing the presently
available neutrino oscillation  experimental 
measurements. 
The SK Collaboration showed that the $\nu_\mu$ created 
in the atmosphere oscillates into $\nu_\tau$ with 
almost maximal mixing  \cite{SK}, $\sin^{2} 2\theta_{atm}\sim 1$
and the neutrino mass squared difference 
$\Delta m^2_{atm}=(2\sim4) \times 10^{-3}~{\rm [eV^2]}$.
The second  mass-squared difference and mixing angle 
are constrained by solar neutrino experiments.
The SNO collaboration reported that the $\nu_e$'s from 
the sun oscillate into the  other active neutrinos \cite{SNO}.
The SK and the SNO collaboration \cite{SNO} suggested 
that the MSW large-mixing-angle (LMA) solution is the most 
favorable solution to the solar-neutrino deficit problem, 
for which $\sin^{2}2\theta_{sol}=0.7\sim 0.9$ and 
the combined analysis of the KamLAND \cite{KM} and 
all the solar neutrino data
gives $\Delta m^2_{sol}=(3\sim 15)\times 10^{-5}~{\rm [eV^2]}$.
For the third mixing angle, only the upper bound is obtained from
the reactor neutrino experiments. CHOOZ \cite{CHOOZ} 
found $\sin^22\theta_{rea} < 0.1$
for $\Delta m^2_{atm} \sim 3\times 10^{-3}~{\rm [eV^2]}$.
The current neutrino experimental data 
indicates clearly that there is a hierarchy of neutrino mass-splitting.
If the small mixing angles $\theta_{L13}$ and $\theta$ are taken,
the light neutrinos mixing matrix $U_{MNS}$ defined in (\ref{MNS}), 
can be simplified,
\bea
U_{MNS}\simeq \left(\begin{array}{ccc}
         \cos\theta_{L12}&\sin\theta_{L12}&\begin{array}{c}
                         \sin\theta_{L13}e^{-i\delta_L}+\\
                         \sin\theta_{L12}\sin\theta e^{-i\phi'}
                          \end{array}\\
        -\sin\theta_{L12}\cos\theta_{L23}&\cos\theta_{L12}
        \cos\theta_{L23}&\sin\theta_{L23}\\
        \sin\theta_{L12}\sin\theta_{L23}
        &-\cos\theta_{L12}\sin\theta_{L23}&\cos\theta_{L23}
         \end{array}\right)P(\alpha'),\nn\\
\label{UMNS}
\eea
where $\phi'=\phi+\gamma_L$, $\alpha'=\alpha-\gamma_L/2$
and $P(\alpha')=diag.(1,e^{i\alpha'},e^{-i\alpha'})$.
A very similar form  to the low energy MNS mixing matrix is  obtained.
In this case, the angles in matrix $U_L$ defined in (\ref{UL}) can be 
related directly to 
the corresponding  neutrino mixing angles and can be determined by 
neutrino experiments. 
In this work, we take the natural hierarchical scenario 
and set the following neutrino masses in (\ref{DIA})
for the corresponding measured neutrino mass differences,   
\bea
n_2=\sqrt{\Delta m^2_{sol}}= 7\times 10^{-3}~\m{[eV]},\ \ 
n_3=\sqrt{\Delta m^2_{atm}}= 5\times 10^{-2}~\m{[eV]}.
\eea
For the angles in $U_L$, we can take,
\bea
\theta_{L12}=\theta_{sol}=\frac{\pi}{6},\ \ 
\theta_{L23}=\theta_{atm}=\frac{\pi}{4}.
\eea
The parameters $x_{1,2}$ defined in (\ref{x-parameters}) 
are constrained by the conditions \cite{our-paper}, 
\bea
|x_1-x_2|\leq n_3-n_2,\ \ n_3+n_2\leq x_1+x_2.
\eea

%\begin{wrapfigure}{r}{6.5cm}
%\begin{wrapfigure}{r}{7.0cm}
%\centerline{\includegraphics[width=6.6cm]{y12.eps}}
%\centerline{\includegraphics[width=6.0cm]{y12.eps}}
%\caption{The evolution of the heavy Majorana Neutrino densities.}
%\label{fig:ynk}
%\end{wrapfigure}

The primordial lepton number must be created when the sphaleron 
process is in equilibrium so that the conversion from the lepton
numbers to the baryon number occurs effectively.
In our numerical
calculation, we fix the mass for the lightest
heavy Majorana neutrino $M_1$ as $2 \times 10^{11}$~[GeV], below 
$T_{\rm sph} \simeq 10^{12}$~[GeV] and we
take the ratio $R=\frac{M_1}{M_2}$ to be $0.1$.
Finally, we consider the constraints from measurements of the 
cosmological baryon to photon ratio $\eta=n_B/n_\gamma 
$\cite{baryon-photon,BOOMERanG}.
Recall that temperature below $T_{\rm EW}$, 
sphaleron process is not active and thus, 
the total baryon number $B$ is conserved.
\begin{wrapfigure}{r}{6.5cm}
\centerline{\includegraphics[width=6.0cm]{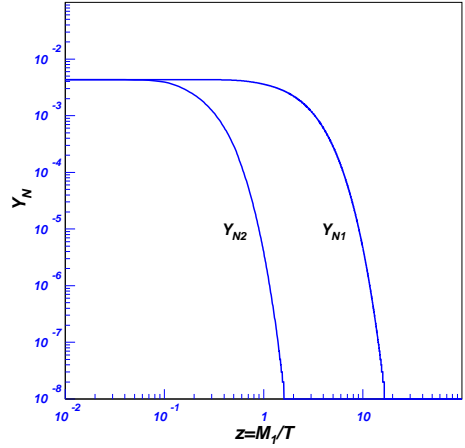}}
\caption{The evolution of the heavy Majorana Neutrino densities.}
\label{fig:ynk}
\end{wrapfigure}
If the universe expands adiabatically,
the total entropy $S$ is also conserved. Then $Y_B$ is also
conserved as,
\bea
Y_B(T_{\rm EW})=Y_B(T_0=3{\rm K})
               \simeq \frac{1}{7}\eta.\nn \\
\label{ybtem}
\eea
Substituting (\ref{ybtem}) into (\ref{conversion}) and combining 
the measurements in (\ref{eta}), we infer the 
bounds on the primordial lepton asymmetry  $Y_L=n_L/s$
that we have to generate,
\bea
&&Y_L(T_{\rm EW})=-\frac{51}{196}\eta \nn \\
               &=&-(1.15-2.13)\times 10^{-10},
\label{ylem}
\eea
at 90\% C.L.  \\
%%%
We take the following values for the other parameters;
\bea
\sin\theta_{L13} e^{-i\delta_L} =0, \quad
(x_1,x_2)=(0.0523,0.0100) ~{\mbox{\rm {[eV]}}}.
\eea
Concerning Higgs mass $m_H$, we change it from $150$(GeV) to $800$(GeV)
and the results are not sensitive to the choice. So we show the figures 
for
$m_H=800~{\mbox{\rm {[GeV]}}}.$
With them, only the CP violating phase $\gamma_L$ is the
parameter which remains to be fixed.

\begin{wrapfigure}{r}{6.5cm}
\centerline{\includegraphics[width=6.0cm]{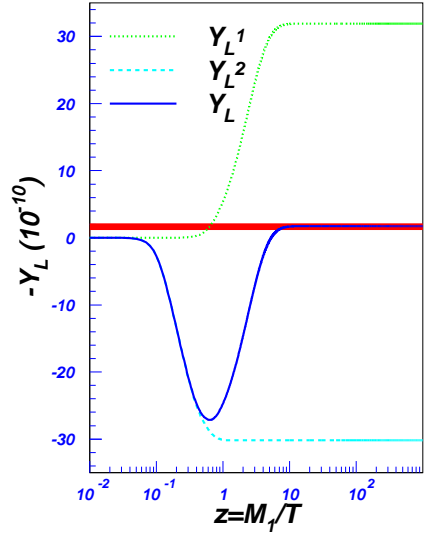}}
\caption{The lepton asymmetry $Y_{L}^1 \ (Y_{L}^2)$ from $N_{R_1}$ 
($N_{R_2}$) decay.
The total lepton asymmetry is denoted by $Y_L$. $\gamma_L=0$.
The shaded part 
shows the
bounds on $Y_L$ in (\ref{ylem}) via experimental measurements 
\cite{baryon-photon,BOOMERanG} at 90\% C.L.}
\label{sum}
\end{wrapfigure}

\begin{figure}[b]
\centerline{\includegraphics[width=11cm]{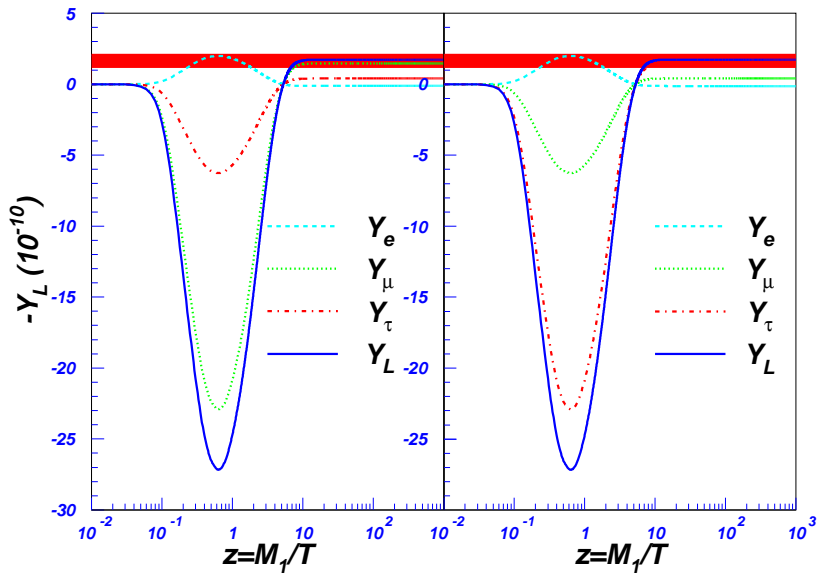}}
\caption{The evolution of the lepton asymmetry and lepton family asymmetries
with the CP violating phase ${\gamma}_{L}=0$ (left) 
and $\gamma_{L}=\pi$ (right).}
\label{yl123-a}
   \end{figure}

\begin{figure}[htb]
\centerline{\includegraphics[width=8cm]{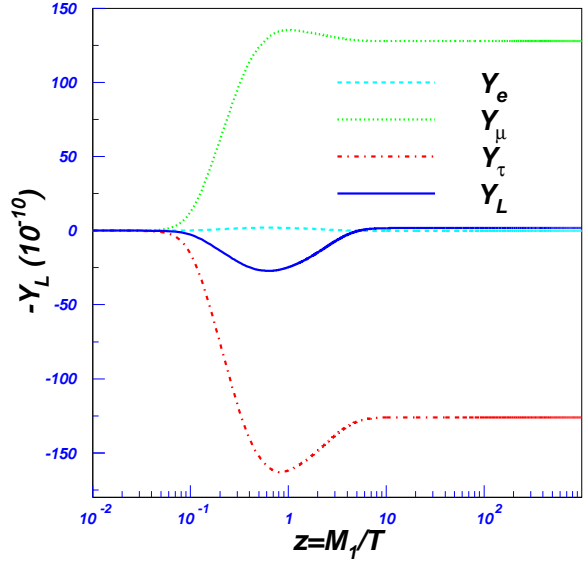}}
\caption{The evolution of the lepton asymmetry and lepton family asymmetries
with the CP violating phase $\gamma_L=\frac{\pi}{2}$.}
\label{yl123-b}
\end{figure}

\begin{figure}[htb]
\centerline{\includegraphics[width=12cm]{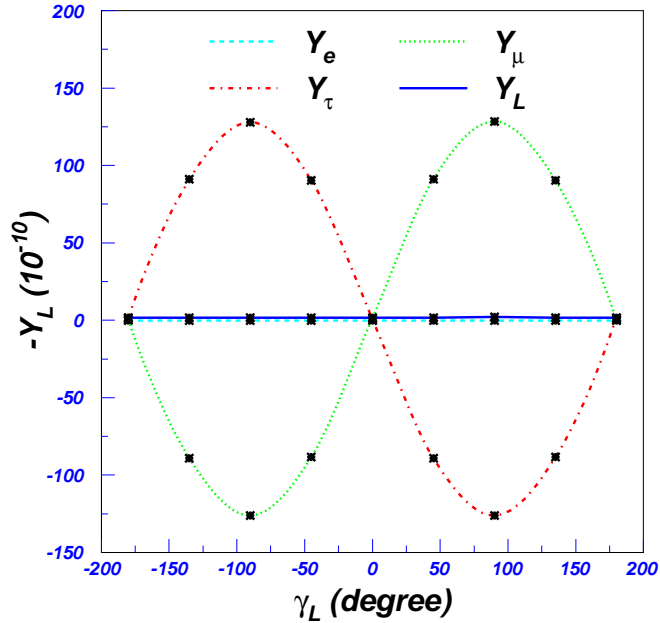}}
\caption{The lepton family asymmetries
as a function of the CP phase $\gamma_L$.
The data points denoted with star are extrapolated.}
\label{yl123-c}
\end{figure}

To solve the Boltzmann equation,
we start at $T \gg M_1$ ($z=10^{-2}$) with the initial conditions,
\bea
Y_{N_k}=Y_{N_k}^{eq},\ \ Y_{L_i}=0.
\eea
The typical results are displayed in Fig.~\ref{fig:ynk} to Fig.~\ref{yl123-c}. 
Fig.~\ref{fig:ynk} shows the dependence of 
the heavy Majorana neutrino density to entropy density ratios 
$Y_N$ on temperature.
One can see clearly that for high temperature region 
$z \leq 0.1$, $Y_{N_1}$ is nearly equal to $Y_{N_2}$. However,
as $z$ becomes larger,  the asymmetry from the heavier 
right-handed neutrino decays drops quickly, thus, the lighter
one gives dominate contribution.

We plot the evolution of the lepton family asymmetries
in Fig.~\ref{sum}, Fig.~\ref{yl123-a} and Fig.~\ref{yl123-b}. 
By investigating the family structure of these 
figures, we observe the
following features.\\
 
%%Their features are given as follows.\\
(1) The total lepton asymmetry $Y_L$ increases with larger 
$z$ and reach its maximum at $z\sim 0.6$. 
The behavior of $Y_L$ in small $z$ region
can be understood as follows. 
Since 
${\rm Im}[(\ynu^\dag\ynu)^2]_{12}=-{\rm Im}[(\ynu^\dag\ynu)^2]_{21}$,
the term $\sum\limits_i\left(\f{Y_{N_k}}{Y_{N_k}^{eq}}-1\right){\epsilon}_{i}^k
{\gamma}_{D}^{ki}$ in (\ref{ynkyli}) 
is positive for $k=2$ while
negative for $k=1$.  In addition, for $R\ll 1$, from Eqs.
(\ref{epsilon}), (\ref{IX}) and  (\ref{gammadki}), one can 
obtain that the order of 
$\sum\limits_i\epsilon_i^k\gamma_{D}^{ki}$ is the same for 
$k=1$ and $k=2$. 
However, around $z \simeq 0.1$, the deviation from equilibrium distribution 
is larger for the heavier Majorana neutrion $N_{R_2}$.
Thus, the magnitude of the positive one
is much larger than the negative one. 
Then, $Y_L$ increases in the small $z$ region.
This feature is in contrast to the previous work
\cite{our-paper} where only contribution from the 
lighter right-handed neutrino $N_{R_1}$ are taken into account.
As $z$ becomes larger, the positive contribution to
$Y_L$ from the heavier
Majorana neutrinos decreases and the contribution
from the lightest heavy Majorana neutrinos
decay becomes dominant. We 
can see that the sign of $Y_L$ changes at the intermediate
regions ($1<z<10$). Then the lepton asymmetry from the
the heavier Majorana neutrino ($N_{R_2}$) is compensated by
the one from the lightest heavy Majorana neutrino ($N_{R_1}$).
This feature can be seen clearly by showing the contribution
to lepton asymmetry
from $N_{R_1}$ and $N_{R_2}$ decays separately, as desplayed
in Fig.~\ref{sum}.
$Y_L$ will further decrease and it tends
to be a constant asymptotically at low temperature.  
It is important to include the contributions from
both heavy Majorana neutrinos for the lepton asymmetry and its
evolution. 

(2) In Fig.~\ref{yl123-a}, we have shown
the lepton family asymmetries for the two different choices
of CP violating phase $\gamma_L$. 
The figure on the left corresponds to
$\gamma_L=0$ and that on the right corresponds to $\gamma_L=\pi$.
Despite of similarity in their shapes, 
the dominant contribution to $Y_L$ comes from $Y_\mu$ 
for $\gamma_L=0$, whereas,  
$Y_\tau$ in  case of  $\gamma_L=\pi$. The results 
indicate the lepton family asymmetries $Y_{L_i}$ are
sensitive to the CP violating phase $\gamma_L$. 
To understand about the feature (2), let us focus attention again 
on term ${\epsilon}_{i}^k{\gamma}_{D}^{ki}$ 
in (\ref{ynkyli}) which is propotental to 
$Im[(\ynu^\dag\ynu)_{12}(y_\nu)^*_{i1}(y_\nu)_{i2}]$.
Since the matrix $\ynu^\dag\ynu$ is not related to the matrix $U_L$,
and  $\gamma_L$ just appears in matrix $U_L$ (\ref{UL}),  
the $\gamma_L$-dependent terms in ${\epsilon}_{i}^k{\gamma}_{D}^{ki}$ 
only come from the quantity  $(y_\nu)^*_{i1}(y_\nu)_{i2}$ in
(\ref{epsilon}).
Using (\ref{UL}) and (\ref{VR}), we otbain   
\bea
(y_\nu)^*_{i1}(y_\nu)_{i2}&=&\frac{2}{v^2}e^{i\gamma_R}
\left\{(m_2^2-m_3^2) \sin \theta_R \cos \theta_R
\left(|(U_L)_{i2}|^2-|(U_L)_{i3}|^2\right)\right.\nn\\
&&\left.+m_2m_3\left[\cos^2\theta_R(U_L)^*_{i2}(U_L)_{i3}
     -\sin^2\theta_R(U_L)_{i2}(U_L)^*_{i3}\right]\right\}.
\label{epsl}
\eea
Note only the second term of (\ref{epsl}) 
is relevant to CP phase $\gamma_L$, and  
is proportional to $(U^*_L)_{i2}(U_L)_{i3}$ given by
\bea
(U^*_L)_{12}(U_L)_{13}&=& 
\frac{1}{2}\sin 2\theta_{L13}\sin\theta_{L12}e^{i({\gamma}_L-{\delta}_{L})}
\simeq 0,\nn\\
(U^*_L)_{22}(U_L)_{23}&=&\frac{1}{2}
\left(-\sin 2\theta_{L13}\sin\theta_{L12}
\sin^2\theta_{L23}e^{-i{\delta}_{L}}\right.\nn\\
     &&\left.+\cos\theta_{L12}\cos\theta_{L13}
\sin 2\theta_{L23}\right)e^{i{\gamma}_L}
\simeq \frac{\sqrt{6}}{8}e^{i{\gamma}_L},\nn\\
(U^*_L)_{32}(U_L)_{33}&=&\frac{1}{2}
\left(-\sin 2\theta_{L13}\sin\theta_{L12}
\cos^2\theta_{L23}e^{-i{\delta}_{L}}\right.\nn\\
     &&\left.-\cos\theta_{L12}\cos\theta_{L13}
\sin 2\theta_{L23}\right)e^{i{\gamma}_L}
\simeq -\frac{\sqrt{6}}{8}e^{i{\gamma}_L}.
\label{UL1}
\eea
From (\ref{epsl}) and (\ref{UL1}) that
in cases of $\gamma_L=0,\pi$, 
the $\gamma_L$-dependent 
terms of ${\epsilon}_{i}^k{\gamma}_{D}^{ki}$ have opposite sign, 
leading to that $Y_\mu$  is dominant in the 
total lepton asymmetry in case of $\gamma_L=0$, 
while $Y_\tau$ is dominant in case of $\gamma_L=\pi$.
In Fig. ~\ref{yl123-b}, we have shown the evolution of the
lepton family asymmetries for another choice, $\gamma_L=\frac{\pi}{2}$.
In this case, $|Y_{\mu}|$ and $|Y_{\tau}|$ are much larger than 
$Y_{L}$. In contrast to the total lepton asymmetry, the lepton 
family asymmetry, for example, $Y_{\mu}$
from $N_{R_2}$ decay is dominant and can not be
upset by the contribution from $N_{R_1}$ decay. 

The understanding is illustrated in Fig.~\ref{yl123-c}.
We show the lepton asymmetries at the low energy and study the
dependence of the asymptotic ($z \gg 1$)
values on the CP violating phase $\gamma_L$.
In contrast to case of $Y_{\mu}$ and $Y_{\tau}$,  
$Y_{e}$ and the total lepton asymmetry $Y_L$ is insensitive to 
the value of CP phase $\gamma_L$ and is nearly constant.  
In addition, we find from this figure that for the inputs we chose, the
size of the lepton family asymmetries $Y_{\mu}$
and $Y_{\tau}$ can be of order  $10^{-8}$ and 
cancel each other mostly, leaving the total lepton 
asymmetry of order  $10^{-10}$ which is consistent with
the experimental observations \cite{baryon-photon,BOOMERanG}.

\section{Conclusions}
\label{sec:con}

In this work we have presented a detailed calculations 
for primordial lepton family asymmetries
in seesaw model by taking into account the effect of 
chemical potentials for all SM particles. 
The  $minimal$ CP violating seesaw model is 
used in obtaining numerical results.
The main results of our paper are as follows.\\
(1) Subject to constraints from 
neutrino  oscillation experiments and observed
cosmological baryon to photon ratio, 
the order of the total lepton asymmetry $Y_L$ can be different
from that for each generations $Y_{L_i}$. 
There exist scenarios in the minimal seesaw model,
both $Y_{\mu}$ and $Y_{\tau}$ are of order $10^{-8}$, 
whereas $Y_L$ is of order $10^{-10}$ due to the cancellation. \\
(2) The lepton family asymmetries
are sensitive to CP violating phase $\gamma_L$ in the matrix $U_L$.
In contrast to this, the dependence of the total lepton  
asymmetry on $\gamma_L$  is quite weak. Note that the total lepton 
asymmetry is sensitive to the CP violating phase $\gamma_R$ in $V_R$
{\cite{our-paper}}.\\
(3) Both of the heavy Majorana neutrinos give
important contributions to 
the primordial lepton family asymmetries.
The heavier Majorana neutrino decay can dominate 
in the lepton family asymmetry.\\
The new features we found provide useful informations and 
may be observable in future  cosmological experiments. 
As a future extension of the research, it would
be interesting to investigate the correlation between CP
violation for the lepton family asymmetries and the low energy
CP violation of neutrino oscillations~\cite{Arafune,Bra,Itow,Yokomakura,Broncano}.
%%%%%%%%%%%%%%%%%%%%%%%%%%%%%%%%%%%%%%%%%
%%%%  Acknowledgements  %%%%%%%%%%%%%%%%%
%%%%%%%%%%%%%%%%%%%%%%%%%%%%%%%%%%%%%%%%% 
\section*{Acknowledgements}

The work of T.M. and Z.X. are supported by the Grant-in-Aid for 
JSPS Fellows (No.1400230400).
We would like to thank M. Pl\"{u}macher, K. Yamamoto, S. K. Kang, S. Kaneko,
M. Tanimoto, A. D. Dolgov, and M. Kakizaki for
useful discussions. 

\appendix
\section{Source terms $C_{l_{i}}$ and $C_{n_{k}}$}

In this Appendix, we show the definitions
of  the source terms $C_{l_{i}}$ and
$C_{n_{k}}$.  
Writing the source term as 
$C_{l_{i}}=C_{l_{i}}^{D}+C_{l_{i}}^{R}$,
$C_{l_{i}}^{D}$ and $C_{l_{i}}^{R}$  
stand for contributions from the decays and the inverse decays 
of the heavy
Majorana neutrinos and two-body scatterings, respectively. 
$C_{l_{i}}^{D}$ is defined as,
\begin{eqnarray}
 C_{l_{i}}^{D}&=& \int d\Pi_{1}d\Pi_{2}d\Pi_{N}(2{\pi})^{4}{\delta}^{4}(p_{1}+p_{2}-p_{N}){\times} \nn \\
& & \sum\limits_{k,s_z}\left\{
f_{N}(s_{z})\left[\left|{\cal M}(N_{R_k}(s_z){\rightarrow}l_{i}^{-}{\phi}^{+})\right|^{2}
(1+f_{{\phi}^{+}})(1-f_{l_{i}^{-}})\right.\right. \nn \\
& & +\left.\left.\left|{\cal M}(N_{R_k}(s_z){\rightarrow}{\nu}_{i}{{\phi}^{0}})\right|^{2}
(1+f_{{\phi}^{0}})(1-f_{{\nu}_{i}})\right.\right. \nn \\
& & -\left.\left.\left|{\cal M}(N_{R_k}(s_z){\rightarrow}l_{i}^{+}
{\phi}^{-})\right|^{2}
(1+f_{{\phi}^{-}})(1-f_{l_{i}^{+}})\right.\right. \nn \\
& & \left.\left.-\left|{\cal M}(N_{R_k}(s_z){\rightarrow}\ov{\nu}_{i}
{{\phi}^{0}}^{*})
\right|^{2}(1+f_{{{\phi}^{0*}}})(1-f_{\ov{\nu}_{i}})\right]\right. \nn \\
& & +\left.(1-f_N(s_z))\left[f_{{\phi}^{-}}f_{l_{i}^{+}}
\left|{\cal M}(l_{i}^{+}{\phi}^{-}{\rightarrow}N_{R_k}(s_z))\right|^{2}
\right.\right. \nn \\
& & \left.\left.+f_{{{\phi}^{0}}^{*}}f_{\ov{\nu}_{i}}\left|{\cal M}
(\ov{\nu}_{i}{{\phi}^{0}}^{*}{\rightarrow}N_{R_k}(s_z))\right|^{2}\right.\right. \nn \\
& &\left.\left. -f_{{{\phi}^{+}}}f_{l^-_{i}}
\left|{\cal M}(l_{i}^{-}{\phi}^{+}{\rightarrow}N_{R_k}(s_z))\right|^{2}\right.\right. \nn \\
& & \left.\left.
-f_{{\nu}_{i}}f_{{\phi}^{0}}\left|{\cal M}({\nu}_{i}{\phi}^{0}{\rightarrow}N_{R_k}(s_z))\right|^{2} \right]\right\},
\label{Cl1}
\end{eqnarray}
where $d\Pi_{i}=\f{d^{3}p_{i}}{(2{\pi})^{3}2E_{i}}$ and $\left|{\cal
M}(a+b+{\cdots}{\rightarrow}i+j+{\cdots})\right|^2$ stands for the
amplitude squared for the process
$a+b+{\cdots}{\rightarrow}i+j+{\cdots}$. 

The contributions from two particles scattering processes can be
divided into four parts,
\begin{equation}
 C_{l_{i}}^{R}=
C_{l_{i}}^{(1)R}+C_{l_{i}}^{(2)R}+C_{l_{i}}^{(3)R}+C_{l_{i}}^{(4)R}.
\end{equation}
The terms $C_{l_{i}}^{(\chi)R}\ (\chi=1,2,3,4)$  are defined as follows,
\bea
C_{l_{i}}^{(1)R}&=&
\int  d\Pi_1d\Pi_2d\Pi_3d\Pi_4(2 \pi)^4 \delta^4 (p_1+p_2-p_3-p_4) 
\times \nn \\
&&\sum_{k,s_z,\alpha}\left[f_N(s_z)f_{\bar{t}^\alpha_R}
|{\cal M}(N_{R_k}(s_{z})\ov{t}_{R}^{\alpha}{\rightarrow}l_{i}^{-}\ov{b}_{L}^{\alpha})|^2(1-\flmi)(1-f_{\bar{b}^\alpha_L})\right.\nn\\
&&\left.+f_{N}(s_{z})f_{b^\alpha_{L}}
|{\cal M}(N_{R_k}(s_z)b^\alpha_{L}{\rightarrow}l_{i}^{-}t^\alpha_{R})|^2
(1-f_{l_{i}^{-}})(1-f_{t^\alpha_{R}})\right.\nn\\
&&\left.-f_N(s_z)f_{t^\alpha_R} |{\cal M}(N_{R_k}(s_z)t_{R}^{\alpha}{\rightarrow}l_{i}^{+}b_{L}^{\alpha})|^2(1-f_{l_{i}^{+}})(1-f_{t_R^\alpha})\right.\nn\\
&&\left.-f_{N}(s_{z})f_{\ov{b}_{L}^\alpha}
|{\cal M}(N_{R_k}(s_z)\ov{b}_{L}^\alpha{\rightarrow}l_{i}^{+}
\ov{t}_{R}^{\alpha})|^2(1-f_{l_{i}^{+}})(1-f_{t_R^\alpha})\right.\nn\\
&&\left.+f_{N}(s_{z})f_{\ov{t}^\alpha_{R}}
|{\cal M}(N_{R_k}(s_{z})\ov{t}^\alpha_{R}
{\rightarrow}{\nu}_{i}\ov{t}^\alpha_{L})|^2
(1-f_{{\nu}_{i}})(1-f_{\ov{t}^\alpha_{L}})\right.\nn\\
&&\left.+f_N(s_z)f_{t^{\alpha}_{L}}|{\cal M}
(N_{R_k}(s_{z}){t}_{L}^{\alpha}{\rightarrow}{\nu}_{i}
t^\alpha_{R})|^2(1-f_{{\nu}_{i}})(1-f_{t_{R}^{\alpha}})\right.\nn\\
&&\left.-f_{N}(s_{z})f_{t_{R}^{\alpha}}|
{\cal M}(N_{R_k}(s_{z})t^\alpha_{R}{\rightarrow}\ov{\nu}_{i}t_{L}^{\alpha})|^2(1-f_{\ov{\nu}_{i}})(1-f_{t_{L}^{\alpha}})\right.\nn\\
&&\left.-f_N(s_z)f_{\ov{t}_{L}^{\alpha}}|{\cal M}(N_{R_k}(s_{z})\ov{t}_{L}^{\alpha}{\rightarrow}\ov{\nu}_{i}\ov{t}_{R}^{\alpha})|^2(1-f_{\ov{\nu}_{i}})
(1-f_{\ov{t}_{R}^\alpha})\right.\nn\\
&&\left.-f_{l_{i}^{-}}f_{\ov{b}_{L}^{\alpha}}
|{\cal M}(l_{i}^-\ov{b}_{L}^{\alpha}{\rightarrow}N_{R_k}(s_{z})\ov{t}_{R}^{\alpha})|^2
(1-f_N(s_{z}))(1-f_{\ov{t}_{R}^{\alpha}})\right.\nn\\
&&\left.-f_{l_{i}^{-}}f_{t_{R}^{\alpha}}|{\cal M}(l_{i}^{-}t_{R}^{\alpha}{\rightarrow}N_{R_k}(s_{z})b_{L}^{\alpha})|^2(1-f_N(s_{z}))(1-f_{b^\alpha_L})\right.\nn\\
&&\left.+f_{l_{i}^{+}}f_{b_{L}^{\alpha}}|{\cal M}(l_{i}^{+}b_{L}^{\alpha}{\rightarrow}N_{R_k}(s_{z})\ov{t}_{R}^{\alpha})|^2(1-f_N(s_z))
(1-f_{\bar{t}^\alpha_R})\right.\nn\\
&&\left.+f_{l_{i}^{+}}f_{\ov{t}_{R}^{\alpha}}
|{\cal M}(l_{i}^{+}\ov{t}^\alpha_{R}{\rightarrow}N_{R_k}(s_z)
\ov{b}_{L}^{\alpha})|^2(1-f_N(s_{z}))(1-f_{\ov{b}_{L}^{\alpha}})\right.\nn\\
&&\left.-f_{{\nu}_{i}}f_{\ov{t}_{L}^{\alpha}}
|{\cal M}({\nu}_{i}\ov{t}^\alpha_{L}{\rightarrow}N_{R_k}(s_{z})\ov{t}_{R}^{\alpha})|^2(1-f_{N}(s_{z}))(1-f_{\bar{t}^\alpha_R})\right.\nn\\
&&\left.-f_{\nu_{i}} f^\alpha_{\bar{t}_R}|{\cal M}({\nu}_{i}t_{R}^{\alpha}{\rightarrow}N_{R_k}(s_{z})t_{L}^{\alpha})|^2(1-f_{N(s_z)})(1-f_{t_{L}^{\alpha}})\right.\nn\\
&&\left.+f_{\ov{\nu}_{i}} f_{t^\alpha_{L}}|{\cal M}
(\ov{\nu}_{i}t_{L}^{\alpha}
{\rightarrow}N_{R_k}(s_{z})t_{R}^{\alpha})|^2(1-f_N(s_z))(1-f_{t^\alpha_R})\right.\nn\\
&&\left.+f_{\ov{\nu}_{i}}f_{\bar{t}^\alpha_{R}}
|{\cal M}(\ov{\nu}_{i}\ov{t}^\alpha_{R}
{\rightarrow}N_{R_k}(s_{z})\ov{t}_{L}^{\alpha})|^2(1-f_{N}(s_{z}))
(1-f_{\ov{t}^\alpha_L})\right],
\label{Cl3}
\eea
where $\alpha$ is color indices of quark.
$C_{l_{i}}^{(\chi)R} (\chi=2,3)$ are given as follows,
\bea
C_{l_{i}}^{(2)R}&=&
\int  d\Pi_1d\Pi_2d\Pi_3d\Pi_4(2 \pi)^4 \delta^4 (p_1+p_2-p_3-p_4) 
\times \nn \\
&&\sum_{k,s_z,\alpha}\left[
f_N(s_z)\flpi|{\cal M}(\Nlptb)|^2
(1-f_{t^\alpha_R})(1-f_{\bar{b}^\alpha_L})\right.\nn\\
&&\left.
+f_N(s_z)\fani|{\cal M}(\Nantt)|^2
(1-f_{\bar{t}^\alpha_L})
(1-f_{t^\alpha_R})\right.\nn\\
&&\left.
-f_N(s_z)\flmi|{\cal M}(\Nlmtb)|^2(1-f_{b^\alpha_L})
(1-f_{\bar{t}^\alpha_R})\right.\nn\\
&&\left.
-f_N(s_z)\fni|{\cal M}(\Nntt)|^2(1-f_{t^\alpha_L})
(1-f_{\bar{t}^\alpha_R})\right.\nn\\
&&\left.
-f_{t^\alpha_R}f_{\bar{b}^\alpha_L}|{\cal M}(\tbNlp)|^2(1-f_N(s_z))(1-\flpi)\right.\nn\\
&&\left.
-f_{\bar{t}^\alpha_L}f_{t^\alpha_R}|
{\cal M}(\ttNan)|^2(1-f_N(s_z))(1-\fani)\right.\nn\\
&&\left.
+f_{b^\alpha_L}f_{\bar{t}^\alpha_R}|{\cal M}(\tbNlm)|^2(1-f_N(s_z))(1-\flmi)\right.\nn\\
&&\left.
+f_{\bar{t}^\alpha_R}f_{t^\alpha_L}
|{\cal M}(\ttNn)|^2(1-f_N(s_z))(1-\fni)
\right].
\label{Cl4}
\eea
\begin{eqnarray}
C_{l_{i}}^{(3)R}&=&
\int  d\Pi_1d\Pi_2d\Pi_3d\Pi_4(2 {\pi})^4{\delta}^4(p_1+p_2-p_3-p_4)
{\times} \nn \\
&&\left\{\sum_j\left[\fpm f_{l_{i}^{+}}{|{\cal M}(l_{i}^{+}{\phi}^{-}{\rightarrow}l_{j}^{-}{\phi}^{+})|^{\prime}}^{2}(1+\fpp)(1-f_{l_{j}^{-}})
\right.\right.\nn\\
&& \left.\left.-\fpp f_{l_{j}^{-}} 
{|{\cal M}(l_{j}^{-}{\phi}^{+}
{\rightarrow}{l}^+_{i}{\phi}^{-})|^{\prime}}^{2}
(1+f_{\phi^-})(1-\flpi)\right.\right. \nn\\
&& \left.\left.+\fpm\flpi |{\cal M}(\lpni)|^{\prime 2}(1+\fpz)(1-\fnj)\right.\right.\nn\\
&&\left.\left.-\fpz\fnj |{\cal M}(\nlpi)|^{\prime 2}(1+\fpm)(1-\flpi)\right.\right.\nn\\
&&\left.\left.+\fpzs\fani |{\cal M}(\anni)|^{\prime 2}(1+\fpz)(1-\fnj)\right.\right.\nn\\
&&\left.\left.-\fpz\fnj |{\cal M}(\nani)|^{\prime 2}(1+\fpzs)(1-\fani)\right.\right.\nn\\
&&\left.\left.+\fpzs\fani |{\cal M}(\anlmj)|^{\prime 2}(1+\fpp)(1-\flmj)\right.\right.\nn\\
&&\left.\left.-\fpp\flmj |{\cal M}(\lmani)|^{\prime 2}(1+\fpzs)(1-\fani)\right.\right.\nn\\
&&\left.\left.+\fpm \flpj |{\cal M}(\lplmj)|^{\prime 2}(1+\fpp)(1-\flmi)\right.\right.\nn\\
&&\left.\left.-\fpp \flmi |{\cal M}(\lmlpj)|^{\prime 2}(1+\fpm)(1-\flpj)\right.\right.\nn\\
&&\left.\left.+\fpm\flpj |{\cal M}(\lpnj)|^{\prime 2}(1+\fpz)(1-\fni)
\right.\right.\nn\\
&&\left.\left.-\fpz\fni |{\cal M}(\nlpj)|^{\prime 2}(1+\fpm)(1-\flpj)
\right.\right.\nn\\
&&\left.\left.+\fpzs\fanj |{\cal M}(\annj)|^{\prime 2}(1+\fpz)(1-\fni)
\right.\right.\nn\\
&&\left.\left.-\fpz\fni |{\cal M}(\nanj)|^{\prime 2}(1+\fpzs)(1-\fanj)
\right.\right.\nn\\
&&\left.\left.+\fpzs\fanj |{\cal M}(\anlmi)|^{\prime 2}(1+\fpp)(1-\flmi)
\right.\right.\nn\\
&&\left.\left.-\fpp\flmi |{\cal M}(\lmanj)|^{\prime 2}(1+\fpzs)(1-\fanj)
\right]\right.\nn\\
&&\left.+\sum_{j\neq i}
\left[\fpm \flpi |{\cal M}(\lplpij)|^{\prime 2}(1+\fpm)(1-\flpj)
\right.\right.\nn\\
&&\left.\left.-\fpm \flpj |{\cal M}(\lplpji)|^{\prime 2}(1+\fpm)(1-\flpi)
\right.\right.\nn\\
&&\left.\left.+\fpm\flpi |{\cal M}(\lpnij)|^{\prime 2}
(1+\fpzs)(1-\fanj)\right.\right.\nn\\
&&\left.\left.-\fpzs\fanj |{\cal M}(\anlpji)|^{\prime 2}
(1+\fpm)(1-\flpi)\right.\right.\nn\\
&&\left.\left.+\fpzs\fani |{\cal M}(\annij)|^{\prime 2}]
(1+\fpzs)(1-\fanj)\right.\right.\nn\\
&&\left.\left.-\fpzs\fanj |{\cal M}(\annji)|^{\prime 2}
(1+\fpzs)(1-\fani)\right.\right.\nn\\
&&\left.\left.+\fpz\fnj |{\cal M}(\nlmji)|^{\prime 2}
(1+\fpp)(1-\flmi)\right.\right.\nn\\
&&\left.\left.-\fpp\flmi |{\cal M}(\lmnij)|^{\prime 2}
(1+\fpz)(1-\fnj)\right.\right.\nn\\
&&\left.\left.+\fpp \flmj |{\cal M}(\lmlmji)|^{\prime 2}
(1+\fpp)(1-\flmi)\right.\right.\nn\\
&&\left.\left.-\fpp \flmi |{\cal M}(\lmlmij)|^{\prime 2}
(1+\fpp)(1-\flmj)\right.\right.\nn\\
&&\left.\left.+\fpzs\fani |{\cal M}(\anlpij)|^{\prime
2}(1+\fpm)(1-\flpj)
\right.\right.\nn\\
&&\left.\left.-\fpm\flpj |{\cal M}(\lpnji)|^{\prime 2}(1+\fpzs)(1-\fani)
\right.\right.\nn\\
&&\left.\left.+\fpz\fnj |{\cal M}(\nnji)|^{\prime 2}(1+\fpz)(1-\fni)
\right.\right.\nn\\
&&\left.\left.-\fpz\fni |{\cal M}(\nnij)|^{\prime 2}(1+\fpz)(1-\fnj)
\right.\right.\nn\\
&&\left.\left.+\fpp\flmj |{\cal M}(\lmnji)|^{\prime 2}(1+\fpz)(1-\fni)
\right.\right.\nn\\
&&\left.\left.-\fpz\fni |{\cal M}(\nlmij)|^{\prime 2}(1+\fpp)(1-\flmj)
\right]\right\},
\label{Cl11}
\end{eqnarray}
where, 
\begin{equation}
|{\cal M}(a+b\to c+d)|^{\prime 2}=
|{\cal M}(a+b\to c+d)|^2-|{\cal M}(a+b\to N_{R_k}\to c+d)|^2.
\end{equation}
Finally, $C_{l_{i}}^{(4)R}$ is given as,
\bea
C_{l_{i}}^{(4)R}&=&
\int  d\Pi_1d\Pi_2d\Pi_3d\Pi_4(2{\pi})^4 \delta^4 (p_1+p_2-p_3-p_4)
\times \nn \\
&&\left\{\frac{1}{2}\sum_j
\left[\fpzs\fpzs |{\cal M}(\psani)|^2(1-\fni)(1-\fnj)\right.\right.\nn\\
&&\left.\left.-\fni\fnj |{\cal M}(\ninj)|^2(1+\fpzs)(1+\fpzs)\right.\right.\nn\\
&&\left.\left.+\fpm\fpm |{\cal M}(\pmlmi)|^2(1-\flmi)(1-\flmj)\right.\right.\nn\\
&&\left.\left.-\flmi\flmj |{\cal M}(\lmlmi)|^2(1+\fpm)(1+\fpm)\right.\right.\nn\\
&&\left.\left.+\fani\fanj |{\cal M}(\anani)|^2(1+\fpz)(1+\fpz)\right.\right.\nn\\
&&\left.\left.-\fpz\fpz |{\cal M}(\pni)|^2(1-\fani)(1-\fanj)\right.\right.\nn\\
&&\left.\left.+\flpi\flpj |{\cal M}(\lplpi)|^2(1+\fpp)(1+\fpp)\right.\right.\nn\\
&&\left.\left.-\fpp\fpp |{\cal M}(\pplpi)|^2(1-\flpi)(1-\flpj)\right.\right]\nn\\
&&+\sum_j\left.\left[\fpm\fpzs |{\cal M}(\pmpzsi)|^2(1-\flmi)(1-\fnj)\right.\right.\nn\\
&&\left.\left.-\flmi\fnj |{\cal M}(\lmni)|^2(1+\fpm)(1+\fpzs)\right.\right.\nn\\
&&\left.\left.+\flpi\fanj |{\cal M}(\lpani)|^2(1+\fpp)(1+\fpz)\right.\right.\nn\\
&&\left.\left.-\fpp\fpz |{\cal M}(\pppzi)|^2(1-\flpi)(1-\fanj)\right.\right.\nn\\
&&\left.\left.+\fpm\fpzs |{\cal M}(\pmpzsj)|^2(1-\flmj)(1-\fni)\right.\right.\nn\\
&&\left.\left.-\flmj\fni |{\cal M}(\lmnj)|^2(1+\fpm)(1+\fpzs)\right.\right.\nn\\
&&\left.\left.+\flpj\fani |{\cal M}(\lpanj)|^2(1+\fpp)(1+\fpz)\right.\right.\nn\\
&&\left.\left.-\fpp\fpz |{\cal M}(\pppzj)|^2(1-\flpj)(1-\fani)\right.\right]\nn\\
&&\left.+\sum\limits_{j\neq i}
\left[\fpp\fpm   |{\cal M}(\phi^+\phi^-\to l_i^-l_j^+)|^2(1-\flmi)(1-\flpj)\right.\right.\nn\\
&&\left.\left.-\flmi\flpj |{\cal M}(l_i^-l_j^+\to \phi^+\phi^-)|^2(1+\fpp)(1+\fpm)\right.\right.\nn\\
&&\left.\left.+\flpi\flmj |{\cal M}(l_i^+l_j^-\to \phi^+\phi^-))|^2(1+\fpp)(1+\fpm)\right.\right.\nn\\
&&\left.\left.-\fpp\fpm |{\cal M}(\phi^+\phi^-\to
l_i^+l_j^-)|^2(1-\flpi)(1-\flmj)\right.\right.\nn\\
&&\left.\left.+\fpz\fpzs |{\cal M}(\phi^0\phi^{0*}\to
\nu_i\bar{\nu}_j)|^2(1-\fni)(1-\fanj)\right.\right.\nn\\
&&\left.\left.
-\fanj\fni |{\cal M}(\nu_i\bar{\nu}_j\to
\phi^0\phi^{0*})|^2(1+\fpz)(1+\fpzs)\right.\right.\nn\\
&&\left.\left.+\fnj\fani |{\cal M}(\nu_j\bar{\nu}_i\to
\phi^0\phi^{0*})|^2(1+\fpz)(1+\fpzs)\right.\right.\nn\\
&&\left.\left.
-\fpz\fpzs |{\cal M}(\phi^0\phi^{0*}\to
\nu_j\bar{\nu}_i)|^2(1-\fani)(1-\fnj)\right.\right.
 \nn \\
&&\left.\left.
+\fpp \fpzs |{\cal M}(\phi^+ \phi^{0*} \to \nu_i l_j^+)|^2
(1-\fni)(1-\flpj) \right.\right. \nn \\
&&\left.\left.-\fni \flpj |{\cal M}(\nu_i l_j^+ \to \phi^+ \phi^{0*})|^2 
(1+\fpp)(1+\fpzs) \right.\right.\nn \\
&& \left.\left.
+\fani \flmj |{\cal M}(\ov{\nu_i} l_j^- \to \phi^- \phi^{0})|^2 
(1+\fpz)(1+\fpm) \right.\right. \nn \\
&& \left.\left.-\fpz \fpm  |{\cal M}(\phi^0 \phi^-
\to \ov{\nu_i} l_j^-)|^2 
(1-\fani)(1-\flmj) \right.\right. \nn \\
&& \left.\left.+\fpz \fpm 
|{\cal M}(\phi^0 \phi^-
\to \ov{\nu_j} l_i^-)|^2 
(1-\flmi)(1-\fanj) 
\right.\right. \nn \\
&& \left.\left.-\flmi \fanj 
|{\cal M}(\ov{\nu_j} l_i^- \to \phi^0 \phi^-)|^2 
(1+\fpm)(1+\fpz) \right.\right. \nn \\
&& \left.\left.+\flpi \fnj
|{\cal M}(\nu_j l_i^+ \to \phi^{0*}\phi^+)|^2 
 (1+\fpp)(1+\fpzs) \right.\right. \nn \\
&& \left.\left.-\fpp \fpzs 
|{\cal M}(\phi^+ \phi^{0*} \to \nu_j l_i^+)|^2 
(1-\fnj)(1-\flpi)
\right]
\right\}.
\label{crij2}
\eea

In deriving (\ref{crij2}), the statistical factors are counted
as follows.
For the first eight processes with $i=j$, we need to multiply
a factor 2 compared with those $i\neq j$ 
since $i=j$ cases corresponds to $|\Delta L_i|=2$ processes.
However, there is a symmetric factor 
$\frac{1}{4}$ in the case of $i=j$, 
whereas $\frac{1}{2}$ for other cases. Therefore we have
a common statistical factor for $i=j$ and $ i\ne j$.

Next we calculate $C_{n_k}$ in (\ref{general}) which can also
divided into decay part $C^D_{n_k}$ and $2\to 2 $ reaction part $C^R_{n_k}$. 
Similar to the calculations of the $C_{l_i}$, 
\bea
C_{n_k}^D&=&
\int d\Pi_1d\Pi_2d\Pi_N(2 \pi)^4 \delta^4 (p_1+p_2-p_N)\times \nn\\
&&\sum_{i,s_z}\left\{ 
(1-f_N(s_z))\left[|{\cal M}(\lpN)|^2\fpm\flpi\right.\right.\nn\\
&&\left.\left.+|{\cal M}(\ansN)|^2\fpzs\fani
+|{\cal M}(\lmN)|^2\fpp\flmi\right.\right.\nn\\
&&\left.\left.+|{\cal M}(\nN)|^2\fpz\fni\right]\right.\nn\\
&&\left.-f_N(s_z)\left[|{\cal M}(\Nn)|^2(1+\fpz)(1-\fni)
\right.\right.\nn \\
&&\left.\left.+|{\cal M}(\Nlm)|^2(1+\fpp)(1-\flmi)\right.\right.\nn \\
&&\left.\left.+|{\cal M}(\Nlp)|^2(1+\fpm)(1-\flpi)\right.\right.\nn \\
&&\left.\left.+|{\cal M}(\Nans)|^2(1+\fpzs)(1-\fani)\right]\right\}.
\label{Cn1}
\eea
Since the smallness of Yukawa couplings of lepton and light quark, only
contributions from the processes involving top (anti-top) quark to 
the reaction part are important, i.e.,
\bea
C_{n_k}^R=C_{n_k}^{(1)R}+C_{n_k}^{(2)R},
\eea
with,
\bea
C_{n_k}^{(1)R}&=&
\int  d\Pi_1d\Pi_2d\Pi_3d\Pi_4(2 \pi)^4 \delta^4 (p_1+p_2-p_3-p_4)
\times \nn \\ 
&&\sum_{i,s_z,\alpha}\left[\flmi f_{\bar{b}^\alpha_L}
|{\cal M}(\lmb)|^2(1-f_N(s_z))(1-f_{\bar{t}^\alpha_R})\right.\nn\\
&&\left.+\flmi f_{t^\alpha_R}|{\cal M}(\lmt)|^2(1-f_N(s_z))
(1-f_{b^\alpha_L})\right.\nn\\
&&\left.+\flpi f_{b^\alpha_L}|{\cal M}(\lpb)|^2(1-f_N(s_z))
(1-f_{t^\alpha_R})\right.\nn\\
&&\left.+\flpi f_{\bar{t}^\alpha_R}|{\cal M}
(\lpt)|^2(1-f_N(s_z))(1-f_{\bar{b}^\alpha_L})\right.\nn\\
&&\left.+\fni f_{t^\alpha_R}|{\cal M}(\ntNt)|^2
(1-f_N(s_z))(1-f_{t^\alpha_L})\right.\nn\\
&&\left.+\fani f_{\bar{t}^\alpha_R}|{\cal M}(\anatNat)|^2
(1-f_N(s_z))(1-f_{\bar{t}^\alpha_L})\right.\nn\\
&&\left.+\fni f_{\bar{t}^\alpha_L}|{\cal M}(\natNat)|^2
(1-f_N(s_z))(1-f_{\bar{t}^\alpha_R})\right.\nn\\
&&\left.+\fani f_{t^\alpha_L}|{\cal M}(\antNt)|^2
(1-f_N(s_z))(1-f_{t^\alpha_R})\right.\nn\\
&&\left.-f_N(s_z)f_{\bar{t}^\alpha_L}
|{\cal M}(\Ntlmb)|^2(1-\flmi)(1-f_{\bar{b}^\alpha_L})\right.\nn\\
&&\left.-f_N(s_z)f_{\bar{b}^\alpha_L}|{\cal M}(\Nblpt)|^2
(1-\flpi)(1-f_{\bar{t}^\alpha_R})\right.\nn\\
&&\left.-f_N(s_z)f_{t^\alpha_L}|{\cal M}(\Ntnt)|^2
(1-\fni)(1-f_{t^\alpha_R})\right.\nn\\
&&\left.-f_N(s_z)f_{\bar{t}^\alpha_L}|{\cal M}(\Natanat)|^2
(1-\fani)(1-f_{\bar{t}^\alpha_R})\right.\nn\\
&&\left.-f_N(s_z)f_{\bar{t}^\alpha_R}|{\cal M}(\Natnat)|^2
(1-\fni)(1-f_{\bar{t}^\alpha_L})\right.\nn\\
&&\left.-f_N(s_z)f_{t^\alpha_R}|{\cal M}(\Ntant)|^2
(1-\fani)(1-f_{t^\alpha_L})\right.\nn\\
&&\left.-f_N(s_z)f_{b^\alpha_L} |{\cal M}(\Nblmt)|^2(1-\flmi)
(1-f^\alpha_{t_R})\right.\nn\\
&&\left.-f_N(s_z)f_{t^\alpha_R}|{\cal M}(\Ntlpb)|^2(1-\flpi)
(1-f_{b^\alpha_L})\right],
\label{Cn11}
\eea
and,
\bea
C_{n_k}^{(2)R}&=&\int  d\Pi_1d\Pi_2d\Pi_3d\Pi_4(2 \pi)^4 \delta^4 (p_1+p_2-p_3-p_4)\times \nn \\
&&\sum_{i,s_z,\alpha}\left[f^\alpha_{t_R}f
_{\bar{b}^\alpha_L}|{\cal M}(\tbNlp)|^2(1-f_N(s_z))(1-\flpi)\right.\nn\\
&&\left.+f_{b_L^{\alpha}}f_{\bar{t}^\alpha_R}|{\cal M}(\tbNlm)|^2(1-f_N(s_z))(1-\flmi)\right.\nn\\
&&\left.
+f_{\bar{t}^\alpha_R}f_{t^\alpha_L}|{\cal M}(\ttNn)|^2(1-f_N(s_z))(1-\fni)\right.\nn\\
&&\left.+f_{\bar{t}^\alpha_L}f_{t^\alpha_R}|
{\cal M}(\ttNan)|^2(1-f_N(s_z))(1-\fani)\right.\nn\\
&&\left.-f_N(s_z)\flpi|{\cal
M}(\Nlptb)|^2(1-f^\alpha_{t_R})(1-f_{\bar{b}^\alpha_L})\right.\nn\\
&&\left.
-f_N(s_z)\flmi|{\cal M}(\Nlmtb)|^2(1-f_{b^\alpha_L})
(1-f_{\bar{t}^\alpha_R})\right.\nn\\
&&\left.
-f_N(s_z)\fni|{\cal M}(\Nntt)|^2(1-f_{\bar{t}^\alpha_L})
(1-f_{t^\alpha_R})\right.\nn\\
&&\left.
-f_N(s_z)\fani|{\cal M}(\Nantt)|^2(1-f_{\bar{t}^\alpha_R})
(1-f_{t^\alpha_L})
\right].
\label{Cn22}
\eea

\section{Reduced cross sections $\hat{\sigma}(\hat{s})$}

%%%%%%%%%%%%%%%%%%%%%%%%%%%%%%%%%%%%%%%%%%%%%%%%%%%%
%%%%%%%%%% < Reduced cross section > %%%%%%%%%%%%%%%
%%%%%%%%%%%%%%%%%%%%%%%%%%%%%%%%%%%%%%%%%%%%%%%%%%%%

The reduced cross sections with $N_{R_k}$ exchange can be expressed as,  
\begin{equation}
 \hat{\sigma}_{N,m}^{ij}=\f{1}{2{\pi}}
\left[\sum_{k}\left|(y_{\nu})_{ik}^{*}(y_{\nu})_{jk}^{*}\right|^{2}
f_{N}^{kk}(\hat{s})
+\sum_{l<k}\Re\left((y_{\nu})_{il}^{*}(y_{\nu})_{jl}^{*}(y_{\nu})_{ik}
(y_{\nu})_{jk}\right)f_{N}^{lk}(\hat{s})\right]
\end{equation}
where $m$ denote the types of the contribution and take
$m=1,2,3,s,t1,t2$ respectively.
$ij$ denote the types of lepton family.
After straightforward calculations, we obtain,
\begin{eqnarray}
 f_{N,1}^{kk}&=& 1+\f{1}{2}\f{\hat{s}a_{k}}{(\hat{s}-a_{k})^{2}+c_{k}a_{k}}+\f{2a_{k}}{D_{k}} \nn \\
&&-\f{a_{k}}{\hat{s}}\left(1+2\f{\hat{s}+a_{k}}{D_{k}}\right)\ln\left(1+\f{\hat{s}}{a_{k}}\right), \\
f_{N,1}^{lk}&=& 2\sqrt{a_{l}a_{k}}\left[\f{\hat{s}}{2D_{l}D_{k}}+\f{1}{D_{l}}+\f{1}{D_{k}}\right. \nn \\
&&+ \left(1+\f{a_{l}}{\hat{s}}\right)\left(\f{1}{a_{k}-a_{l}}-\f{1}{D_{k}}\right)\ln\left(1+\f{\hat{s}}{a_{l}}\right) \nn \\
&&+ \left.\left(1+\f{a_{k}}{\hat{s}}\right)\left(\f{1}{a_{l}-a_{k}}-\f{1}{D_{l}}\right)\ln\left(1+\f{\hat{s}}{a_{k}}\right)\right], \\
 f_{N,2}^{kk}&=& \f{2\hat{s}}{\hat{s}+a_{k}}+\f{4a_{k}}{\hat{s}+2a_{k}}\ln\left(1+\f{\hat{s}}{a_{k}}\right), \\
f_{N,2}^{lk}&=& \f{4\sqrt{a_{l}a_{k}}}{(a_{l}-a_{k})(\hat{s}+a_{l}+a_{k})}\left[(\hat{s}+2a_{l})\ln\left(1+\f{\hat{s}}{a_{k}}\right)\right. \nn \\
&& \left.-(\hat{s}+2a_{k})\ln\left(1+\f{\hat{s}}{a_{l}}\right)\right],\\
f_{N,3}^{kk}&=&\f{1}{2}\f{\hat{s}a_{k}}{(\hat{s}-a_{k})^{2}+c_{k}a_{k}},\\
f_{N,3}^{lk}&=&\sqrt{a_{l}a_{k}} \f{\hat{s}}{D_{l}D_{k}},\\
 f_{N,s}^{kk}&=&\f{1}{2}\f{\hat{s}^{2}}{(\hat{s}-a_{k})^{2}+c_{k}a_{k}},~~~f_{N,s}^{lk}=\f{\hat{s}^{2}}{D_{l}D_{k}},\\
 f_{N,t1}^{kk}&=&\f{\hat{s}}{\hat{s}+a_{k}},~~~f_{N,t1}^{lk}=-\f{2\sqrt{a_{l}a_{k}}}{a_{l}-a_{k}}\left[\ln\left(1+\f{\hat{s}}{a_{l}}\right)-\ln\left(1+\f{\hat{s}}{a_{k}}\right)\right],\\
f_{N,t2}^{kk}&=&-2+\left(1+2\f{a_{k}}{\hat{s}}\right)
\ln\left(1+\f{\hat{s}}{a_{k}}\right),\\
f_{N,t2}^{lk}&=&-2+\f{2}{a_{l}-a_{k}}\left[a_l\left(1+\f{a_{l}}{\hat{s}}
\right)
\ln\left(1+\f{\hat{s}}{a_{l}}\right)\right.\nn\\
&&\left.-a_k\left(1+\f{a_{k}}{\hat{s}}\right)
\ln\left(1+\f{\hat{s}}{a_{k}}\right)\right].
\eea
The corresponding reduced cross sections are defined by,
\bea
\hat{\sigma}_{N,1}^{ij}&\equiv &4\hat{\sigma}(\hat{s})(\lplmi),\ \
\hat{\sigma}_{N,2}^{ij}\equiv 4\hat{\sigma}(\hat{s})(\lmlmi),\\
\hat{\sigma}^{ij}_{N,3}&\equiv &4\hat{\sigma}(\hat{s})(l_i^+ \phi^-
\rightarrow \nu_j \phi^0),\ \ 
\hat{\sigma}^{ij}_{N,s}\equiv 4\hat{\sigma}(\hat{s})(\lplpij), \\
\hat{\sigma}^{ij}_{N,t1}&\equiv& 4\hat{\sigma}(\hat{s})(\lmni),\ \   
\hat{\sigma}^{ij}_{N,t2}\equiv 4\hat{\sigma}(\hat{s})
(l_i^+l_j^-\to \phi^+\phi^-). 
\eea

For the s channel Higgs exchange process
$N_{R_k}l_{i}^{-}{\rightarrow}\ov{t}^{\alpha}b^{\alpha}$, the reduced
cross section reads,
\begin{equation}
 \hat{\sigma}_{{\phi},s}^{ki}\equiv 4\sum_{\alpha}
\hat{\sigma}(\hat{s})(N_{R_k}l^-_i\to \bar{t}^\alpha b^\alpha)
=\f{N_{c}}{4{\pi}}\left(y_{t}^{\dagger}y_{t}\right)\left|(y_{\nu})_{ik}\right|^{2}\left(1-\f{a_{k}}{\hat{s}}\right)^{2},
\label{sigma-phis}
\end{equation}
where $N_{c}$ is the color number of quark , $y_t$ is top quark Yukawa coupling 
, and for the t channel Higgs exchange
process $N_{R_k}\ov{t}^{\alpha}{\rightarrow}l_{i}^{-}\ov{b}^{\alpha}$,
\begin{eqnarray}
\hat{\sigma}^{ki}_{\phi,t}&\equiv& 
4\sum_{\alpha}\hat{\sigma}(\hat{s})(N_{R_k}\bar{t}^\alpha\to l^-_i\bar{b}^\alpha)
=\f{N_{c}}{4{\pi}}\left(y_{t}^{\dagger}y_{t}\right)\left|(y_{\nu})_{ik}\right|^{2}\left(1-\f{a_{k}}{\hat{s}}\right) \nn \\
& &{\times} \left[\f{\hat{s}-2a_{k}+2a_{H}}{\hat{s}-a_{k}+a_{H}}-\f{a_{k}-2a_{H}}{\hat{s}-a_{k}}\ln\f{a_{H}}{\hat{s}-a_{k}+a_{H}}\right],
\label{sigma-phit}
\end{eqnarray}
where $a_{k},\ \hat{s}$ are defined in (\ref{nkeq})  and (\ref{resc}),
respectively, $a_{H}=M_{H}^{2}/{M_{1}^{2}}$ with $m_H$ being the Higgs
mass, $c_{k}=\left({\Gamma_D^{k}}/{M_{1}}\right)^{2},\ \Gamma_D^{k}=4\sum\limits_i\Gamma_{ki}$,
\begin{equation}
 \f{1}{D_{k}}=\f{\hat{s}-a_{k}}{(\hat{s}-a_{k})^{2}+a_{k}c_{k}},
\end{equation}
is the off-shell propagator of $N_{R_{k}}$.

%%%%%%%%%%%%%%%%%%%%%%%%%%%%%%%%%%%%%%%%%%%%%%%%%%%%%
%%%%%%%%%%%%%%%%% <References> %%%%%%%%%%%%%%%%%%%%%%%
%%%%%%%%%%%%%%%%%%%%%%%%%%%%%%%%%%%%%%%%%%%%%%%%%%%%%
 
\end{document}